\documentclass[conference]{IEEEtran}
\IEEEoverridecommandlockouts
\usepackage{cite}
\usepackage{amsmath,amssymb,amsfonts}
\usepackage{algorithm}
\usepackage{graphicx}
\usepackage{textcomp}
\usepackage{xcolor}
\usepackage{glossaries}
\usepackage{siunitx}
\usepackage[caption=false]{subfig}
\usepackage[noend]{algpseudocode} 
\usepackage{gensymb} 

\def\BibTeX{{\rm B\kern-.05em{\sc i\kern-.025em b}\kern-.08em
    T\kern-.1667em\lower.7ex\hbox{E}\kern-.125emX}}

\newacronym{VR}{VR}{Virtual Reality}
\newacronym[plural=HMDs]{HMD}{HMD}{Head-Mounted Display}
\newacronym{mmWave}{mmWave}{millimeter-wave}
\newacronym{AP}{AP}{Access Point}
\newacronym{(MU-)MIMO}{(MU-)MIMO}{(Multi-User) Multiple Input Multiple Output}
\newacronym{ULA}{ULA}{Uniform Linear Array}
\newacronym{URA}{URA}{Uniform Rectangular Array}
\newacronym{Slerp}{Slerp}{Spherical linear interpolation}
\newacronym[plural=UAVs, longplural=Unmanned Aerial Vehicles]{UAV}{UAV}{Unmanned Aerial Vehicles}
\newacronym{COTS}{COTS}{Commercial Off-The Shelf}
\newacronym[plural=AoAs]{AoA}{AoA}{Angle of Arrival}
\newacronym{SNR}{SNR}{Signal-to-Noise Ratio}
\newacronym{LoS}{LoS}{Line-of-Sight}
\newacronym{EIRP}{EIRP}{Effective Isotropic Radiated Power}
\newacronym{AWV}{AWV}{Antenna Weight Vector}
\newacronym{MCS}{MCS}{Modulation and Coding Scheme}

\DeclareSIUnit{\belmilliwatt}{Bm}
\DeclareSIUnit{\belisotropic}{Bi}
\DeclareSIUnit{\dBm}{\deci\belmilliwatt}
\DeclareSIUnit{\dBi}{\deci\belisotropic}

\newcommand*{\transpose}{^{\mathrm{T}}}
\newcommand*{\compconj}{^{\mathrm{*}}}

\MakeRobust{\Call} 
\algnewcommand\algorithmicto{..}
\algrenewtext{For}[3]{\algorithmicfor\ #1 $\gets$ #2\algorithmicto#3 \algorithmicdo}

\begin{document}

\title{Millimeter-Wave Beamforming with Continuous Coverage for Mobile Interactive Virtual Reality
}

\author{\IEEEauthorblockN{Jakob Struye, Filip Lemic and Jeroen Famaey}
\IEEEauthorblockA{IDLab - Department of Computer Science\\
University of Antwerp - imec, Antwerp, Belgium\\
Email: \{jakob.struye,filip.lemic,jeroen.famaey\}@uantwerpen.be}
}

\maketitle

\begin{abstract}
Contemporary Virtual Reality (VR) setups commonly consist of a Head-Mounted Display (HMD) tethered to a content-generating server. "Cutting the wire" in such setups and going truly wireless will require a wireless network capable of delivering enormous amounts of video data at an extremely low latency. Higher frequencies, such as the millimeter-wave (mmWave) band, can support these requirements. Due to high attenuation and path loss in the mmWave frequencies, beamforming is essential. For VR setups, beamforming must adapt in real-time to the user's head rotations, but can rely on the HMD's built-in sensors providing accurate orientation estimates. In this work, we present coVRage, a beamforming solution tailored for VR HMDs. Based on past and current head orientations, the HMD predicts how the Angle of Arrival (AoA) from the access point will change in the near future, and covers this AoA trajectory with a dynamically shaped beam, synthesized using sub-arrays. We show that this solution can cover such trajectories with consistently high gain, unlike regular single-beam solutions.
\end{abstract}

\section{Introduction}
A wide variety of \gls{VR} applications have been investigated over the years, in fields including education, medicine and manufacturing~\cite{VRapplications1,VRapplications2,VRapplications3}. Such applications require a reliable high-throughput and low-latency connection to an external device providing \gls{VR} content~\cite{VRChallenges}. This may be live video recorded elsewhere, such as for remote collaboration or meetings, or 3D graphics generated on a PC or edge cloud, such as for gaming applications. The recently introduced Oculus Quest 2 \gls{HMD} is capable of generating content on-device and thereby working without any connection to other devices. As this restricts the device from running many such connected and computationally intensive applications, it also offers the option to tether it to a PC. This setup, along with most others currently on the market, relies on a wired connection for content delivery. While this easily meets reliability, latency and throughput requirements, it limits the user's range of movement, hindering true immersion.\\To achieve truly wireless connected \glspl{HMD}, \gls{mmWave} networking, in the \SI{30}{} to \SI{300}{\giga\hertz} band, is most often considered, as lower frequencies cannot meet the \gls{VR} requirements~\cite{VRChallenges}. Solutions often rely on the existing IEEE 802.11ad and IEEE 802.11ay Wi-Fi standards for \gls{mmWave}~\cite{PerasoVR} or on 5G NR's \gls{mmWave} capabilities~\cite{VrMecFallback}. 
The main challenges in building such a system stem from \gls{mmWave}'s inherently high path loss and attenuation. To achieve sufficiently high signal strength at the \gls{HMD}, the transmitter and the \gls{HMD} must both focus their energy towards each other, in a process called beamforming. \Gls{mmWave} transceivers usually implement beamforming using phased antenna arrays, consisting of many separate antenna elements~\cite{Fundamentals}. The path lengths of the signal from each element will differ slightly in a given direction, meaning the different signals are generally not phase-aligned. By carefully shifting the phase of each element, a beamforming algorithm ensures signals towards an intended receiver are phase-aligned and therefore at maximum amplitude. As this phenomenon also applies to signals received at phased arrays, beamforming should also occur when in receive mode, focusing towards the transmitter. While basic beamforming consists of a single beam in one direction, we exploit a more advanced approach using a variable number of sub-beams. By subdividing the array into sub-arrays providing sub-beams, the combined beam can cover a dynamically shaped area.\\
Such flexible coverage is highly advantageous for beamforming on an \gls{HMD}. An angular beam misalignment of a few degrees can have a significant impact on \gls{SNR}~\cite{MoVR}, and a human head can reach an instantaneous angular velocity of hundreds of degrees per second~\cite{OScan,RotationDataset1,RotationDataset2}. As such, a flexibly shaped beam, stretched in the direction of a head rotation, can provide an \gls{HMD} with consistently high receive gain, essential for uninterrupted low-latency video delivery. To form such a beam proactively, head rotations must be accurately predicted. Fortunately, \glspl{HMD} are, by design, equipped with orientation estimation capabilities. Current and historical estimations enable the design of reasonably accurate predictors of future orientations and rotations. \\
In this paper, we present \textit{coVRage}, a novel beamforming method for \glspl{HMD}, supporting uninterrupted connectivity during rapid head movements. This is, to the best of our knowledge, the first \gls{HMD}-focused beamforming method offering proactive \gls{AoA} trajectory coverage through sub-arrays. Using simulation, we demonstrate that coVRage provides a stable gain in a single-user \gls{VR} scenario.\\
The remainder of this paper is organised as follows. In Section~\ref{sec:rw}, we provide background and related work on sub-arrays, \gls{mmWave} \gls{VR} and head rotation prediction. Section~\ref{sec:antenna} investigates how phased arrays may be placed within an \gls{HMD}, along with an appropriate system model. Section~\ref{sec:orientations} outlines how to represent 3D orientations. Next, Section~\ref{sec:algo} presents coVRage, and in Section~\ref{sec:eval} we evaluate how well it performs in simulation. Finally, Section~\ref{sec:conclusion} concludes this paper.
\section{Background and Related Work} \label{sec:rw}
\subsection{Sub-arrays}
\begin{figure}[!t]
    \centering
    \begin{minipage}{1.0\linewidth}
    \subfloat[Localized sub-arrays]{\includegraphics[width=0.45\textwidth]{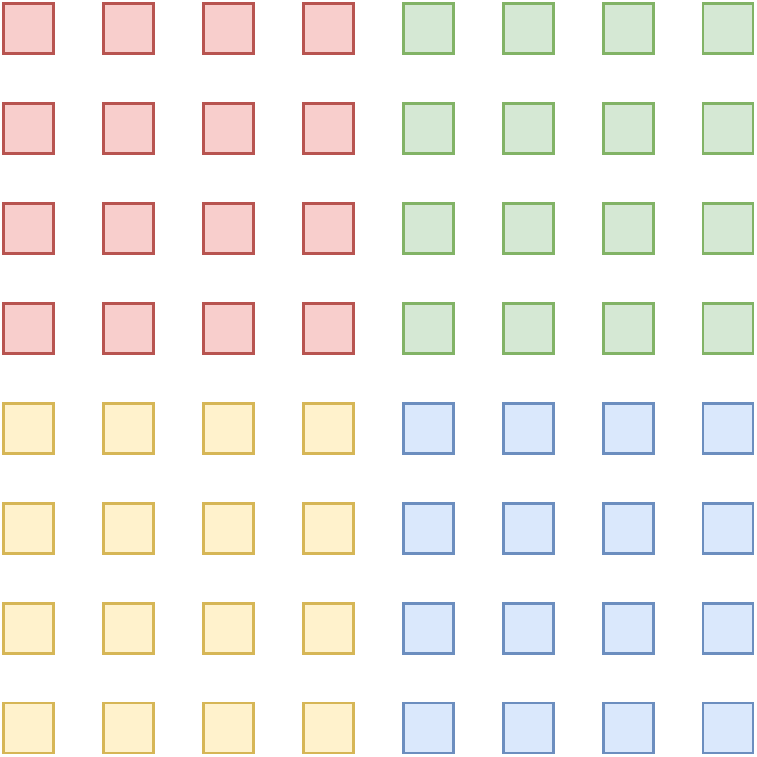}}
    \hfill
    \subfloat[Interleaved sub-arrays]{\includegraphics[width=0.45\textwidth]{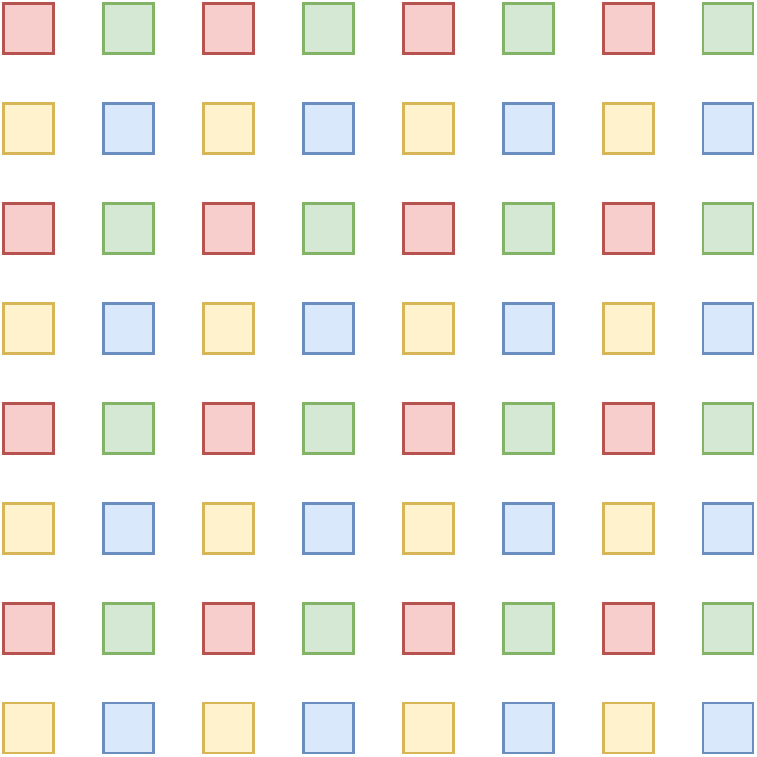}\label{fig:interleaved}}
    \end{minipage}
    \caption{Localized and interleaved sub-arrays in a Uniform Rectangular Array}
    \label{fig:layouts}
\end{figure}

To form beams of flexible size and shape, sub-arrays are crucial. Therefore, we provide an overview of sub-arrays and the related literature. A sub-array may be either localized, with all elements adjacent, or interleaved, with elements spread across the entire array, as illustrated in Fig.~\ref{fig:layouts}. The sub-array configuration can be supported at a hardware level, by having multiple RF chains, allowing each sub-array to send a different signal. This includes hybrid arrays, with one chain per sub-array, and digital arrays, with one chain per element~\cite{ShimuraInterleaved}. When only one RF chain is available for all elements, the array is called analog. Several works present design decisions for hybrid and digital arrays for localized~\cite{subarr,Multimode,EnergyEfficient,ChannelEst} or interleaved~\cite{ShimuraInterleaved, interleaved5G} sub-array antennas. 
Zhang \emph{et al.} compare the two in terms of performance and feasibility~\cite{LocalizedvsInterleaved}. 
For beamforming with hybrid arrays, many approaches have been proposed. These may either form a single main lobe~\cite{HybridArray_Huang,HybridAdaptive}, or provide simultaneous coverage for multiple users~\cite{HybridSDMA,HybridMultiplex, LowComplexMulticast,AnalogBeamforming}.
The hybrid phased array has also been used to design hierarchical codebooks, facilitating a binary-search approach to beamforming with gradually narrowing beams~\cite{HybridHierarchical}. Physical sub-arrays, based on the array's design, can be further subdivided into logical sub-arrays. This allows for more flexible hierarchical codebook design~\cite{PartiallyHybridCodebook}. Such codebooks can also be designed with logical sub-arrays only, which only requires an analog array~\cite{VirtualHierarchicalOld, VirtualHierarchical,AnalogHierarchicalMimo}. Multi-user coverage with logical sub-arrays has also been investigated, both by assigning a sub-array per user~\cite{AnalogMultiuser}, or by synthesizing one large beam of flexible shape, covering all recipients~\cite{FlexibleCoverage}. Our algorithm extends this final approach to cover the upcoming trajectory of one peer, rather than the current locations of several peers.
\subsection{Wireless \gls{VR}}
Several works have considered \gls{mmWave} for cutting the cord in \gls{VR}. In the MoVR solution, a ceiling-mounted relay assists the \gls{AP} at the edge of the playing field~\cite{MoVR}. The \gls{HMD}'s built-in location and orientation tracking is used to steer transmit and receive beams directly at peers. Zhong \emph{et al.} present a programmable \gls{mmWave} wireless solution using \gls{COTS} hardware and investigate rendering-based optimizations~\cite{cotsMMVR}. Other works further investigate such optimizations~\cite{renderingVR,OffloadingVR}. 
Elbamby \emph{et al.} outline the challenges of \gls{mmWave} \gls{VR}~\cite{VRChallenges}. Na \emph{et al.} measure attainable \gls{VR} throughput with \gls{COTS} IEEE 802.11ad hardware~\cite{PerasoVR}. The IEEE 802.11ad standard was shown to be a good fit for interactive \gls{VR}, with its channel access settings having a significant impact on the attainable datarate~\cite{struye2020towards}. Kim, Lee and Lee propose a dynamic power control algorithm for energy-efficient \gls{VR} delivery over IEEE 802.11ad~\cite{EnergyEfficientVR}. Several proposed designs incorporate falling back to legacy Wi-Fi to cover \gls{mmWave} signal loss~\cite{ProactiveFallback,VrMecFallback}. 
In case of pre-recorded \gls{VR} content, frames can be sent proactively over \gls{mmWave} using predicted future viewing directions~\cite{ProactiveTransmission}. Pose information-assisted networks leverage location and orientation measurements from on-device sensors, such as in \glspl{HMD}, for beam selection as well as \gls{AP} selection, focused on spatial sharing between clients~\cite{Pia}. Finally, OScan proposes fast 3D beam steering for mobile clients such as \glspl{HMD}, using UV-coordinates~\cite{OScan}. Of these works, only OScan considers \gls{HMD}-side beamforming, but it does not support proactively covering upcoming \glspl{AoA}. As such, our work is complementary to most of the aforementioned works.
\subsection{Head Rotation Prediction}
Several approaches of varying complexity have been considered for head rotation prediction. A variety of works has shown the effectiveness of classical approaches such as autoregression and Kalman filters for head rotation estimation and prediction~\cite{KF5, KF3, KF4,DeltaQ}. The more recent field of viewport prediction essentially solves the same problem~\cite{Viewport1,Viewport2}. Recent work uses deep learning to further improve the results~\cite{ViewportDL1}, and may use video content as additional inputs~\cite{ViewportDL2}. While the different approaches are difficult to compare directly due to varying prediction horizons and datasets, most approaches provide predictions amply accurate for our application. Several of the above approaches achieve an average error under a third of that of a baseline predictor which outputs the latest known orientation as prediction.
\section{Assumptions and System Model} \label{sec:antenna}
In this section, we describe the environment coVRage is expected to operate in, provide array design guidelines based on this environment, and outline an appropriate system model.
\subsection{Expected Environment}
CoVRage considers a \gls{VR} setup where a ceiling-mounted \gls{mmWave} \gls{AP} serves an \gls{HMD}-wearing user on the ground. The user can freely rotate their head. Within the time span of a single rotation, the user's location is expected to remain static (the location intuitively changes more slowly than the rotation). The \gls{AP} is assumed to run some beamforming algorithm enabling it to always perfectly focus its beam at the \gls{HMD}. The \gls{HMD} can estimate its own orientation with high accuracy, and can accurately predict its orientation in the near future~\cite{riftAccuracy,RiftTracking}. Given this orientation, the \gls{HMD} is able to derive the direction towards the \gls{AP}. The \gls{HMD} is equipped with a \gls{mmWave} phased array. The goal of coVRage is then to tune the receive beam of the \gls{HMD} such that the received signal strength is consistently high while the \gls{HMD} rotates towards the predicted orientation. CoVRage achieves this by synthesizing a beam covering the entire (shortest) trajectory between the current and predicted orientation. The prediction horizon should be large enough to encompass a single fast head movement, e.g., \SI{200}{\milli\second}.
\subsection{Antenna Array Design}
The antenna array for the \gls{HMD} should be designed with the expected environment outlined above in mind. We provide some guidelines, then present a specific design.\\First of all, we eliminate hybrid and digital arrays. While their many RF chains would offer more flexible beamforming, their power consumption and cost are prohibitive for a battery-powered consumer device~\cite{VirtualHierarchical}. We therefore opt for an analog array. A next trade-off to consider is between the number of elements in the array, and the spacing between these elements. For an $N$-element \gls{ULA}, the attainable beamwidth in radians is
\begin{equation}
    b_\alpha = \frac{0.886 \lambda}{Nd\cos\alpha}\label{eq:bw}
\end{equation}
at a steering angle $\alpha$ ($\alpha=0$ being broadside), with an inter-element spacing of $d$. Such a \gls{ULA}, with all elements on one line, will however not suffice, as it can only beamform with one degree of freedom~\cite{phaseBook}. As coVRage requires 3D beamforming, with both azimuth and elevation of the beam controllable, a \gls{URA} is needed. For a \gls{URA} of size $N=N_xN_y$ aimed at $(\phi,\theta)$, the azimuthal and elevational beamwidths are calculated separately, replacing $N$ and $\alpha$ in \eqref{eq:bw} with either $N_x$ and $\phi$ or with $N_y$ and $\theta$. The beamwidth equation implies that, for a fixed physical area, adding more elements within said area will not tighten the beamwidth. As such, an inter-element spacing of $d=0.5\lambda$ is often used throughout the industry, as a tighter spacing leads to unwieldily wide beams, while wider spacing is known to create \textit{grating lobes}; undesired side lobes with a directional gain as high as the main lobe's. This rule of thumb, however, no longer applies when using interleaved sub-arrays. With $M_i$ interleaved sub-arrays in a \gls{URA}, the inter-element spacing within the sub-array is is $\sqrt{M_i}d$, as illustrated by Fig.~\ref{fig:interleaved}. As such, the physical inter-element spacing should be chosen with a specific $M_i$ in mind. Whenever the sub-beams that these $M_i$ interleaved sub-arrays can create are unable to cover a full trajectory, they should be further subdivided into \textit{sub-sub-arrays}, which would be localized within the sub-array.\\
For the remainder of this paper, we consider a specific instantiation of the phased array within the \gls{HMD}. Measuring many modern \glspl{HMD} showed that a square \gls{URA} of length \SI{4}{\centi\metre} is feasible. We will use the \SI{60}{\giga\hertz} band, as this unlicensed band is free to use, and already widely used for \gls{mmWave} Wi-Fi. Then, we use $M_i = 4$ interleaved sub-arrays, meaning the inter-element spacing becomes $d = 0.25\lambda =$ \SI{0.125}{\centi\metre}. At this configuration, creating sub-sub-arrays would lead to rather large beams, meaning this is mainly a feasible option for higher frequencies. At \SI{300}{\giga\hertz}, often considered the upper limit of \gls{mmWave}, a sub-sub-array could consist of $40\times40$ elements, having a beamwidth of only \SI{2.54}{\degree}.
\subsection{System Model}\label{sec:system}
CoVRage is a receiver-side beamforming method, which assumes a \gls{LoS} path always exists and ignores reflected paths\footnote{With the indoor ceiling-to-floor transmissions we consider, \gls{LoS} is unlikely to be broken. First-order reflections are most likely via walls, and their power is assumed to be negligible as long as the user is not right next to the wall and grating lobes are avoided. Redirected walking~\cite{RedirectedWalking} can keep mobile users away from walls.}. As such, we opt for a simple system model~\cite{EnergyEfficientVR,MOCA,ZeroOverhead}, calculating the received power as
\begin{equation}
    P_R = P_T + G_T - PL(d) + G_R \label{eq:pr}
\end{equation}
where $P_T$ and $P_R$ are input and received power in \SI{}{\dBm}, $G_T$ and $G_R$ are transmitter and receiver gain in \SI{}{\dBi} and $PL(d)$ is the path loss over $d$ meters in \SI{}{\deci\bel}.\\
Transmitter-side beamforming is assumed to be perfect\footnote{If pose information is forwarded from \gls{HMD} to \gls{AP}, beamforming at the (static) \gls{AP} is considerably simpler than at the (rotating) \gls{HMD}, and therefore considered to be solved for the purpose of our channel model.}, so the transmitter's \gls{EIRP} is constantly at the maximum legally allowed strength (\SI{30}{\dBm} in Europe), and
\begin{figure}[!t]
    \centering
    \begin{minipage}{1.0\linewidth}
    \subfloat[ULA path]{\includegraphics[width=0.485\textwidth]{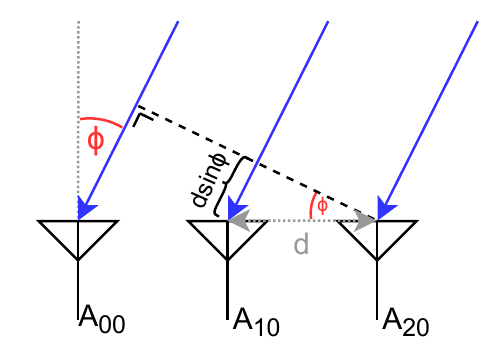} \label{fig:ula}}
    \hfill
    \subfloat[URA path]{\includegraphics[width=0.485\textwidth]{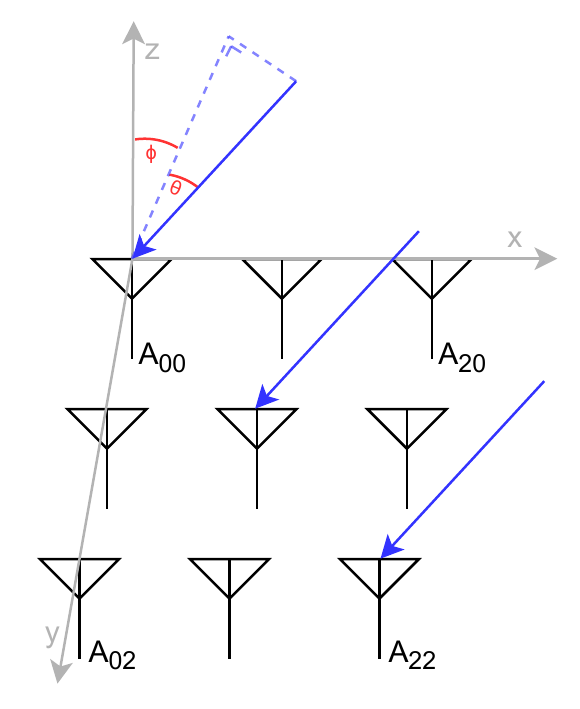} \label{fig:ura}} 
    \end{minipage}
    \caption{At an \gls{AoA} $\phi$, the path shortens by $d\sin\phi$ for every next element for a \gls{ULA}. With a \gls{URA} and \gls{AoA} $(\phi,\theta)$, this becomes $d\sin\phi\cos\theta$ in the $x$-direction and $d\sin\theta$ in the $y$-direction. }
    \label{fig:antenna}
\end{figure}
\begin{equation}
    EIRP = P_T + G_T
\end{equation}
Using the well-known log-distance path loss model, we approximate the path loss as 
\begin{equation}
    PL(d) = PL_{fs}(d_0) + 10n \log_{10}\left(\frac{d}{d_0}\right)
\end{equation}
where $d$ is the transmitter-receiver distance in meters, $d_0$ is some reference distance, $n$ is the path loss exponent and $PL_{fs}(d_0)$ is the Friis free-space path loss over $d_0$:
\begin{equation}
    PL_{fs}(d_0) = 20\log_{10}\left(\frac{4\pi d_0}{\lambda}\right)
\end{equation}
where $\lambda$ is the wavelength. The path loss exponent is estimated as $2$ for an indoor \gls{LoS} \gls{mmWave} setting~\cite{pathloss}, so given a wavelength of \SI{0.005}{\metre} for \SI{60}{\giga\hertz}, and using $d_{0}=\SI{1}{\metre}$, the model simplifies to approximately
\begin{equation}
    PL(d) = 68 + 20\log_{10}(d)
\end{equation}
To determine the receiver gain, we first determine the phase shift between antenna elements. For a \gls{URA} of size $N_xN_y$, using element $A_{0,0}$ as reference element, the phase shift becomes~\cite{phaseBook}
\begin{equation}
    \delta_{x,y}(\phi,\theta) = e^{j 2 \pi d \lambda^{-1} (-x\sin\phi \cos\theta - y\sin \theta)}
\end{equation}
for element $A_{x,y}$ with an \gls{AoA} of azimuth $\phi$ and elevation $\theta$, as illustrated in Fig.~\ref{fig:antenna}.
Then, the phase shifters of the receive array are configured with \gls{AWV} $\mathbf{w}$ with $N_xN_y$ complex elements each with magnitude 1, such that the received signal is modified with coefficient
\begin{equation} \label{eq:coeff}
    C_R(\phi,\theta) = \sum_{x=0}^{N_x-1}\sum_{y=0}^{N_y-1} [\mathbf{w}]_{x, y} \delta_{x,y}(\phi,\theta)
\end{equation}
such that the final directional receive gain in \SI{}{\dBi} for some \gls{AWV} and \gls{AoA} is
\begin{equation}
    G_R(\phi,\theta) = 10\log_{10}(|C_R(\phi,\theta)|^2) \label{eq:gain}
\end{equation}
where $(\phi,\theta)$ may be omitted for brevity when they represent the AoA.\\
To beamform the receiver towards a specific direction, its gain must be maximized. For this, the weight elements $[\mathbf{w}]_{x, y}$ of weight $\mathbf{w}$ must be set to:
\begin{equation}\label{eq:beamform}
    [\mathbf{w}]_{x, y} = \frac{1}{\delta_{x,y}} = e^{j 2 \pi d \lambda^{-1} (x\sin\phi \cos\theta + y\sin \theta)}
\end{equation}
\section{Orientations and directions} \label{sec:orientations}
Several methods of representing orientations and directions in 3D space have seen common use over the years~\cite{Angles1,Angles2,Angles3}. Each has its own advantages and disadvantages, meaning no single most useful representation exists, and care must be taken to select the most appropriate representation for an application. These representations may vary in interpretability, compactness, uniqueness, numerical stability, computational efficiency, ease of combination/subdivision and susceptibility to gimbal lock. Different graphical \gls{VR} engines supply user orientations in different representations, and throughout coVRage several representations are deliberately used to exploit their advantages.
\subsection{Representations}
An easily interpretable representation is that of the \textbf{Euler angles}. In this system, an orientation is described by three chained rotations around the three axes of the coordinate system, where this coordinate system rotates along with the body. As 3D rotations are not commutative, the order of orientations must be properly defined.
The separate rotations are often referred to as \textit{yaw}, \textit{pitch} and \textit{roll}, assigned the variables $\phi$, $\theta$ and $\psi$ respectively.
This is easily converted from an orientation to a direction; by simply omitting the final rotation, a direction in 3D space is represented compactly. In this interpretation, the two remaining rotations are frequently called the \textit{azimuth} and \textit{elevation}.\\
In graphical engines, rotations are often represented by \textbf{unit quaternions}. Quaternions, first covered in the mid 19th century, are an extension of complex numbers, containing three imaginary units $i$, $j$ and $k$, all equal to $-1$ when squared, rather than just the one. In this paper, we represent the quaternion $w+xi+yj+zk$ as the vector $\mathbf{q} = [w, x, y, z]\transpose$. The set of unit quaternions (i.e., of norm 1) is a double-cover of the 3D rotation group, meaning that for each rotation in 3D space, exactly two unit quaternion representations exist ($\mathbf{q}$ and $-\mathbf{q}$, as negating both the magnitude and axis of a rotation results in the same rotation). Quaternions are mathematically convenient; they are numerically stable, do not suffer from gimbal lock and are computationally efficient. Furthermore, quaternions are easily combined by simply multiplying them using the Hamilton product. When representing a vector $\vec{v}$ as quaternion $\mathbf{v} = [0, \vec{v}_x, \vec{v}_y, \vec{v}_z]\transpose$, the product $\mathbf{v}' = \mathbf{q}\mathbf{v}\mathbf{q}\compconj$, where $\mathbf{q}\compconj$ is the complex conjugate, represents $\vec{v}$ rotated by $\mathbf{q}$. Interpolation and extrapolation are also simple: $\mathbf{q}^a$ maintains the rotational axis but multiplies the magnitude by $a$.\\
As a final representation, we consider \textbf{uv-coordinates}~\cite{UVcoords, OScan}. $(u,v)$, consisting of only two real variables, only has enough degrees of freedom to represent directions in 3D, similar to the azimuth-elevation representation. UV-coordinates however exist in sine-space, meaning $\{(u,v) | u,v \in [-1,1]\}$ covers a hemisphere whose center $(0,0)$ is equivalent to azimuth and elevation 0.
Why these coordinates are commonly used for beamforming is outlined in Section \ref{sec:algo2}.

\subsection{Conversions} \label{sec:conversions}
As different components within the beamforming system presented in this paper require different representations of orientations and directions, we often need to convert between them. The following conversions are used for the remainder of the paper.
\subsubsection{Quaternions to Euler angles}
To convert a  quaternion $\mathbf{q} = [w, x, y, z]\transpose$ to Euler angles $(\phi, \theta, \psi)$, calculate~\cite{Angles1}:
\begin{equation}\label{eq:quattoeuler}
    \begin{aligned}
    \phi &= \arctan \frac{2(wx+yz)}{1 - 2(x^2+y^2)}\\
    \theta &= \arcsin (2(wy+xz))\\
    \psi &= \arctan \frac{2(wz+xy)}{1 - 2(y^2+z^2)}
    \end{aligned}
\end{equation}
where the arctangent must be implemented using the \texttt{atan2} function, returning a result in $[-\pi,\pi]$.
\subsubsection{Euler angles to UV-coordinates} 
For this conversion, first convert the orientation to a direction, by simply discarding the roll $\psi$. Then, the UV-coordinates are~\cite{OScan}
\begin{equation}\label{eq:eulertouv}
    \begin{aligned}
        u=&\cos\theta\sin\phi\\
        v=&\sin\theta
    \end{aligned}
\end{equation}

Note that this definition differs from the one commonly used for the similar UV-mapping in graphical engines, which covers the full sphere.
\subsubsection{UV-coordinates to Euler angles} 
In the opposite direction, $\phi$ and $\theta$ can be recovered as
\begin{equation}\label{eq:uvtoeuler}
    \begin{aligned}
    \phi &= \arctan \frac{u}{\sqrt{1 - u^2 - v^2}}\\
    \theta &= \arcsin v
    \end{aligned}
\end{equation}
again using \texttt{atan2} in the implementation. This clearly shows that not every $(u,v)$ is a valid coordinate. If $u^2 + v^2 > 1$, the azimuth is no longer a real number, meaning such coordinates are invalid.
\section{CoVRage} \label{sec:algo}
In this section, we provide a step-by-step explanation of how coVRage works, along with a brief analysis of its computational efficiency.

\subsection{The Algorithm}
CoVRage must convert measured current and predicted future \gls{HMD} orientations to a set of phase shifts for the phased array in the \gls{HMD}. We decompose this process into three distinct steps. First, we determine how the \gls{AP} appears to move relative to the \gls{HMD}, the reference point. Specifically, we determine the direction of the \gls{AP} at the start and end of the rotation between \gls{HMD} orientations, and the shortest trajectory between these directions, in UV-space. Next, we determine a set of beams that covers this trajectory, achievable by the phased array. Finally, we minimize the destructive interference between the sub-arrays on the trajectory, to avoid having "blind" spots along the trajectory.
\begin{figure}[!t]
    \centering
    \includegraphics[width=\linewidth]{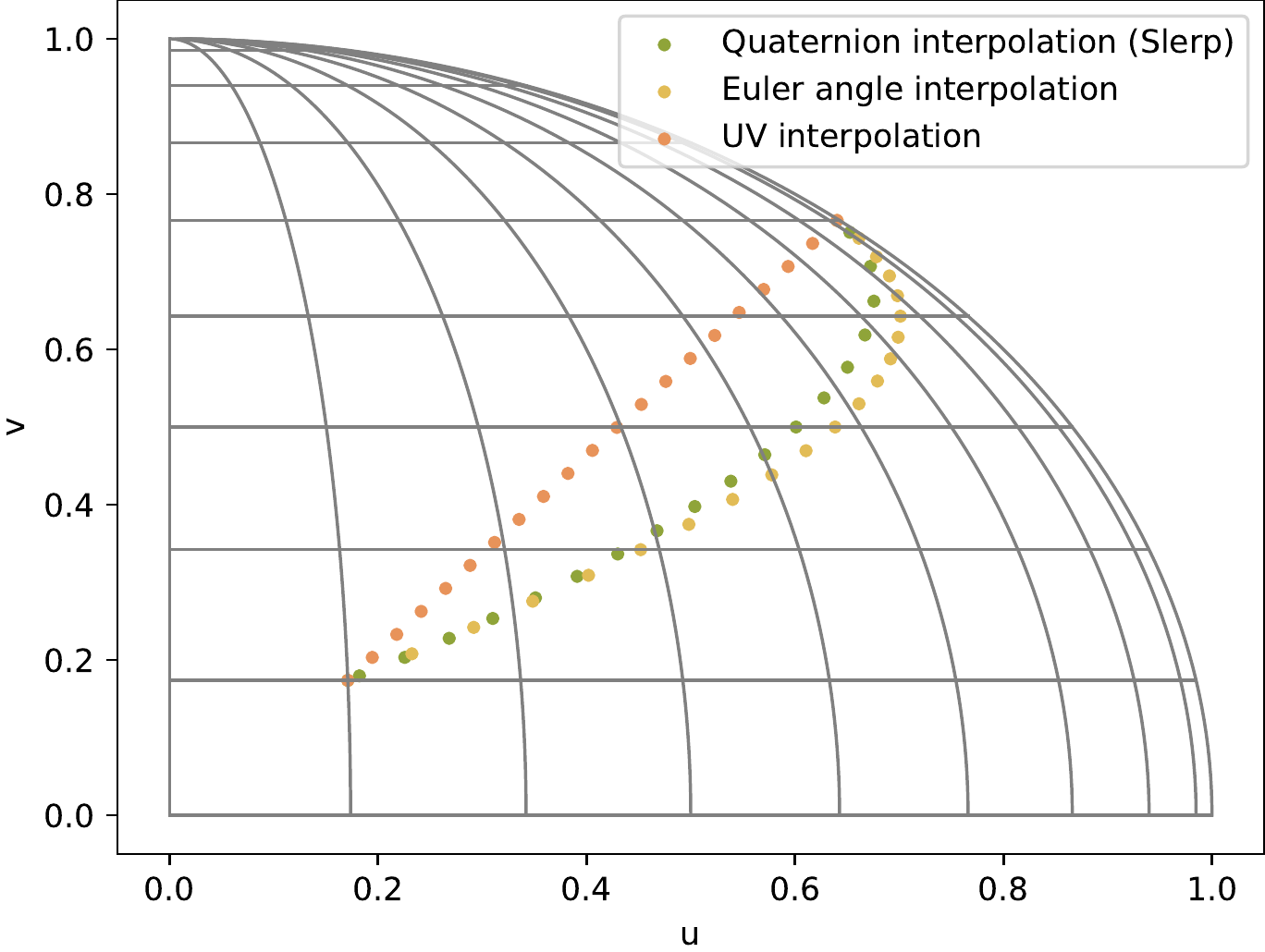}
    \caption{Interpolated path between $(10\degree,10\degree)$ and $(85\degree,50\degree)$, performed with different representations. Only Slerp shows the actual shortest path.}
    \label{fig:interp}
\end{figure}
\subsubsection{Trajectory Generation}\label{sec:beamtraj}
To present trajectory generation, we borrow some terminology from 3D graphics. All objects in 3D space are located relative to the world coordinate system, which is attached to the \gls{HMD}. During a head rotation, this world coordinate system rotates. It is simple to see that this is equivalent to applying the inverse rotation, around the world coordinate system, to all other objects in 3D space. 
In quaternion terms, the \gls{HMD} rotates from orientation $\mathbf{q}_1$ to $\mathbf{q}_2$ by rotation $\mathbf{q}_2\mathbf{q}_1\compconj$, meaning the \gls{AP} will appear to perform the rotation $(\mathbf{q}_2\mathbf{q}_1\compconj)\compconj = \mathbf{q}_1\mathbf{q}_2\compconj$ around the user. To translate rotations to absolute directions, the \gls{AP} direction at one point must be known. This can be hard-coded, or measured using existing \gls{AP} sensing approaches~\cite{Pia}.\\
As the \gls{HMD} is only expected to provide the start and end of the expected rotation within some brief time-frame, coVRage is responsible for generating the path of the \gls{AP} direction during the rotation, between those two points. The representation of the orientation depends on the used framework. OpenVR provides rotation matrices, Unreal uses Euler angles and Unity gives quaternions. The goal of this step was to determine the \gls{AP} trajectory in UV-space, so some conversion is definitely required. Furthermore, determining the shortest trajectory between two orientations (i.e., a single rotation, known to exist from Euler's rotation theorem) is not straightforward with UV-coordinates. To generate this UV-space trajectory, we will need to first generate it in another representation, sample some points from it, and convert those to UV-coordinates. More directly, we need to \textit{interpolate} between the points. In the 3D graphics world, it is widely known that naive interpolation does not work well with rotation matrices and Euler angles~\cite{Slerp}, as this does not generate orientations on the shortest trajectory between the reference orientations. Quaternions, on the other hand, are known to be a perfect fit for interpolation. Given two quaternions $\mathbf{q}$ and $\mathbf{p}$, the quaternion $\mathbf{p}\mathbf{q}\compconj$ represents the rotation from the orientation represented by $\mathbf{q}$ to that represented by $\mathbf{p}$. The set of quaternions $(\mathbf{p}\mathbf{q}\compconj)^a$ for $a \in [0,1]$ covers exactly all intermediate orientations achieved during said rotation. This algorithm is known as \gls{Slerp} and widely used in 3D graphics~\cite{Slerp}. The resulting quaternions are easily converted to UV-coordinates using \eqref{eq:quattoeuler} and \eqref{eq:eulertouv}. 
As trajectories have only a modest curve in UV-space, this approximation is very close. Fig.~\ref{fig:interp} shows interpolations performed with quaternions, Euler angles and UV-coordinates. As only the quaternion-based interpolation provides the shortest path, this is used in coVRage.
\subsubsection{Sub-Beamforming}\label{sec:algo2}
Once the \gls{AP} trajectory as seen from the \gls{HMD} is determined, the algorithm needs to synthesize a beam covering it. As the beam will consist of a variable number of sub-beams from sub-arrays, the number of beams, and, as an effect, their width, must first be determined. Here, the choice for UV-coordinates becomes clear. Remember from \eqref{eq:bw} that the beamwidth depends on the angular distance from broadside ($\alpha=0$). In UV-coordinates however, the beamwidth is nearly invariant to the beam's direction~\cite{UVcoords}. As such, the beamwidth in UV-space can be approximated by the constant
\begin{equation}
    b_{uv} = \frac{0.886 \lambda}{Nd}
\end{equation}
\begin{figure}[!t]
    \centering
    \begin{minipage}{1.0\linewidth}
    \subfloat[Euler Angles]{\includegraphics[width=0.485\textwidth]{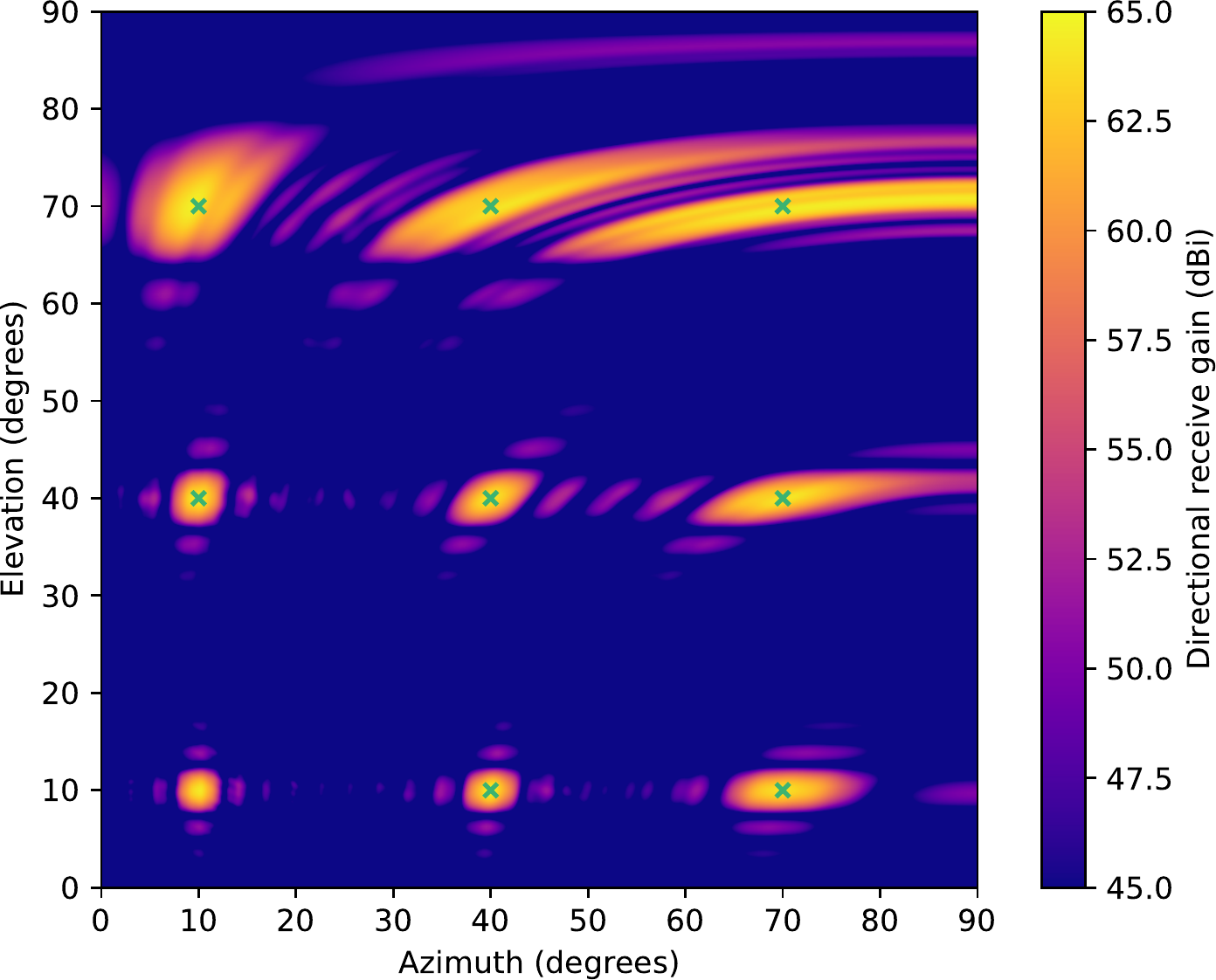} \label{fig:spread1}}
    \hfill
    \subfloat[UV-coordinates]{\includegraphics[width=0.485\textwidth]{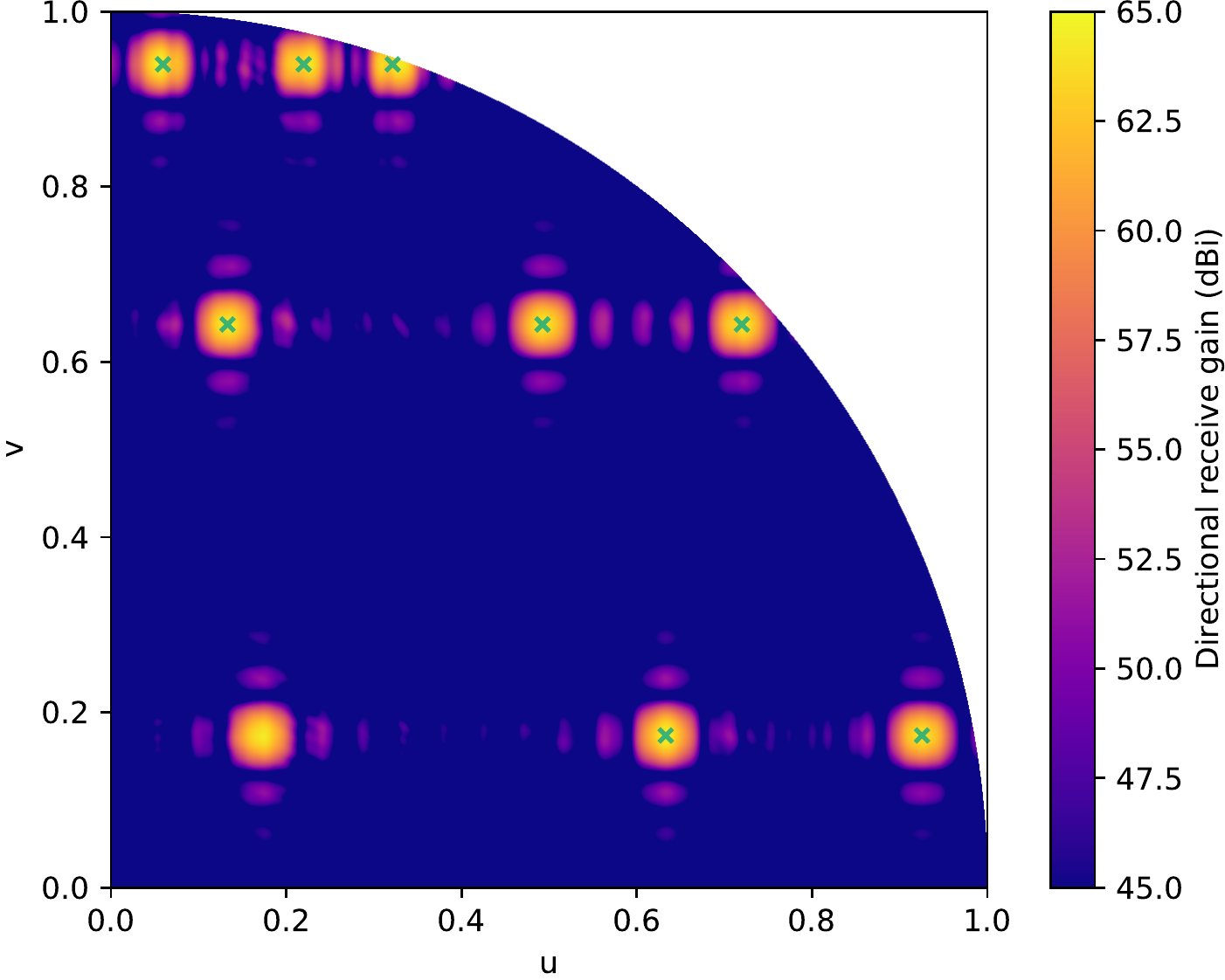} \label{fig:spread2}} 
    \end{minipage}
    \caption{9 beams, equally spaced in Euler angle-space, all appear as near-perfect circles in UV-space.}
    \label{fig:spread}
\end{figure}
with an error always under \SI{2}{\percent}, highest near the edges of the hemisphere. As shown in Fig.~\ref{fig:spread}, a rectangular sub-array's beam anywhere in UV-space is as such accurately represented by a circle of constant radius, eliminating the need for complicated, time-consuming beam shape calculations. The problem of trajectory coverage with sub-beams is essentially reduced to covering a curve  using circles. The first substep here is to determine how many beams are needed to cover the entire trajectory, noting that more beams means fewer elements per beam and therefore wider beams. Estimate the trajectory length $l_t$ as the sum of distances between adjacent UV-space trajectory points. We will aim the first sub-beam towards the current direction, then divide the remaining beams along the trajectory such that each point lies within at least one sub-beam's beamwidth. When each interleaved sub-array has a beamwidth of $w_i$ (0.111 for $16\times16$), the required number of sub-arrays $M_s$ is
\begin{equation}
    M_s = \left\lceil \frac{l_t + 0.5w_i}{w_i} \right\rceil
\end{equation}
as only half the beamwidth of the first sub-beam, aimed at the first point, covers the trajectory. Experimentation showed that aiming the first sub-beam such that the first point is at the edge of its beamwidth provided insufficient coverage at that first point.\\
If $M_s$ exceeds the available number of interleaved sub-arrays $M_i$ (4 for the \SI{4}{\centi\metre} $\times$ \SI{4}{\centi\metre} array), each must be further subdivided into localized sub-sub-arrays. For each subdivision, each sub-beam's width doubles and the number of sub-beams quadruples, meaning the required number of subdivisions is the minimal value of $s$ for which
\begin{equation}
    l_t + 2^{(s-1)} w_i  <= 4^sM_i\,2^sw_i
\end{equation}
As the coverable trajectory length at \SI{60}{\giga\hertz} is already over $3$ for $s=1$, this is mainly of practical use with higher frequencies.\\
Another possibility is that fewer than the available number of interleaved sub-arrays are needed. Some approaches choose to simply deactivate unneeded sub-arrays~\cite{VirtualHierarchical}, which requires hardware support. We instead reinforce sub-beams by steering multiple sub-arrays in the same direction. When only one sub-array is needed, all are aimed in the same direction, effectively eliminating the sub-arraying mechanism entirely. With two sub-beams required, diagonally located pairs of sub-arrays steer towards the same direction. Finally, when three of the four are needed, the first sub-beam, closest to the current \gls{AP} direction, is formed by two diagonally located sub-arrays.\\
Once the number of beams is determined, aiming these is relatively straightforward. CoVRage iterates through the available sample points on the trajectory curve and determines for each point if a beam should be aimed towards it. This is determined by checking if a sub-beam focused at the point would cover all previously considered points not yet covered by a previous sub-beam. As long as this is the case, no candidate sub-beams are locked in. However, once a candidate sub-beam could no longer cover all as of yet uncovered previous points, the candidate sub-beam at the \textit{previous} point is selected. To avoid coverage gaps between two adjacent sample points, we may require that a sub-beam also covers the most recently considered point already covered by the previously selected sub-beam. As such, two consecutive sub-beams will overlap at (at least) one sample point. With this algorithm, a sub-beam covering the final part of the trajectory may not be found. If this occurs, we extrapolate the trajectory and continue the algorithm until all original sample points are covered. The current implementation uses a simple linear extrapolation using the final two sample points. Using this set of sub-beams, phase shift weights for sub-beam syncing can be calculated. Experiments showed that the impact of how sub-beams are mapped to sub-arrays is negligible.\\
\begin{figure}[!htp]
    \centering
    \begin{algorithm}[H]
        \caption{The coVRage algorithm}
        \label{alg:bf}
        \begin{algorithmic}

    \Function{Subdivide}{$t_l$}
        \State $c \leftarrow (M_i-0.5)0.886 / N$
        \State $s \leftarrow 0$
        \While {$c < t_l$}
            \State $s \leftarrow s+1$
            \State $c \leftarrow 8c$
        \EndWhile
        \State $A \leftarrow \Call{DoInterleavedSubArrays}{M_i, N}$
        \For{i}{1}{s}
            \State $A \leftarrow \Call{DoLocalizedSubArrays}{4, N/i, A}$
        \EndFor
        \State $N_s = N / (s+1)$ \Comment{No. of els per sub-arr. in 1 dir.}
        \State \textbf{return} $A$ \Comment{List of sub-arrays}
    \EndFunction
    \State
    \Function{CoverPoints}{$P,A$}
    \State $B \leftarrow [\ ]$ \Comment{Beams}
    \State $M \leftarrow [\ ]$ \Comment{Points in P covered by 2 adjacent beams}
    \State $p_u \leftarrow $Null \Comment{The earliest not yet covered point}
    \State $p_b \leftarrow P[0]$ \Comment{Most recently allocated beam}
    \State $B \leftarrow \Call{ListAdd}{B,\Call{AimBeam}{p_b, A, N_s}}$

    \ForAll {$p \in P$}
    \State /*Does $N_s$-sized arr. aimed at $p$ cover $p_u$?*/
    \If{$\textbf{not } \Call{Covers}{p_u,p,N_s}$}
    \State $p_b \leftarrow p_{prev}$
    \State $B \leftarrow \Call{ListAdd}{B,\Call{AimBeam}{p_b, A, N_s}}$
    \State $M \leftarrow \Call{ListAdd}{M, p_u}$
    \State $p_u \leftarrow$ Null
    \EndIf
    \If{$p_u$ \textbf{is} Null $\textbf{and not } \Call{Covers}{p,p_b,N_s}$}
        \State $p_u \leftarrow p_{prev}$
    \EndIf
    \State /*At final point but end of trajectory uncovered?*/
    \If{$\Call{IsFinal}{P,p} \textbf{ and } p_u \textbf{ is not}$ Null}
        \State $P \leftarrow \Call{Append}{P,\Call{Extrap}{p_{prev},p}}$
    \EndIf
    \State $p_{prev} \leftarrow p$
    \EndFor
    \State \textbf{return} $B, M$
    \EndFunction
    \State
    \Function{PhaseSyncBeams}{$B,M$}
    \State $B_{sync} \leftarrow B[0]$
    \For{i}{0}{\Call{Len}{M}-2}
        \State $m \leftarrow m[i]$
        \State $b_1 \leftarrow B[i]$, $b_2 \leftarrow B[i+1]$
        \State $\delta_\phi \leftarrow \Call{GetPhase}{b_2, m}-\Call{GetPhase}{b_1, m}$
        \State $b_2 \leftarrow \Call{DoPhaseShift}{b_2, -\delta_\phi}$
        \State $\Call{ListAdd}{B_{sync},b_2}$
    \EndFor
    \State \textbf{return} $B_{sync}$
    \EndFunction
    \State
    \State \textbf{Input:} Trajectory points $P$ with traj. length $t_l$
    \State $N \leftarrow 16$ \Comment{Number of elements in 1 direction, default}
    \State $M_i \leftarrow 4$ \Comment{Number of interleaved sub-arrays, default}
    \State $A \leftarrow \Call{Subdivide}{t_l}$
    \State $B,M \leftarrow \Call{CoverPoints}{P,A}$
    \State $B \leftarrow \Call{PhaseSyncBeams}{B,M}$
    \State \textbf{Output:} $B$
    \end{algorithmic}
\end{algorithm}
\end{figure}
\begin{figure*}[!t]
    \centering
    \begin{minipage}{1.0\linewidth}
    \subfloat[Trajectory A, Euler angles]{\includegraphics[width=0.32\textwidth]{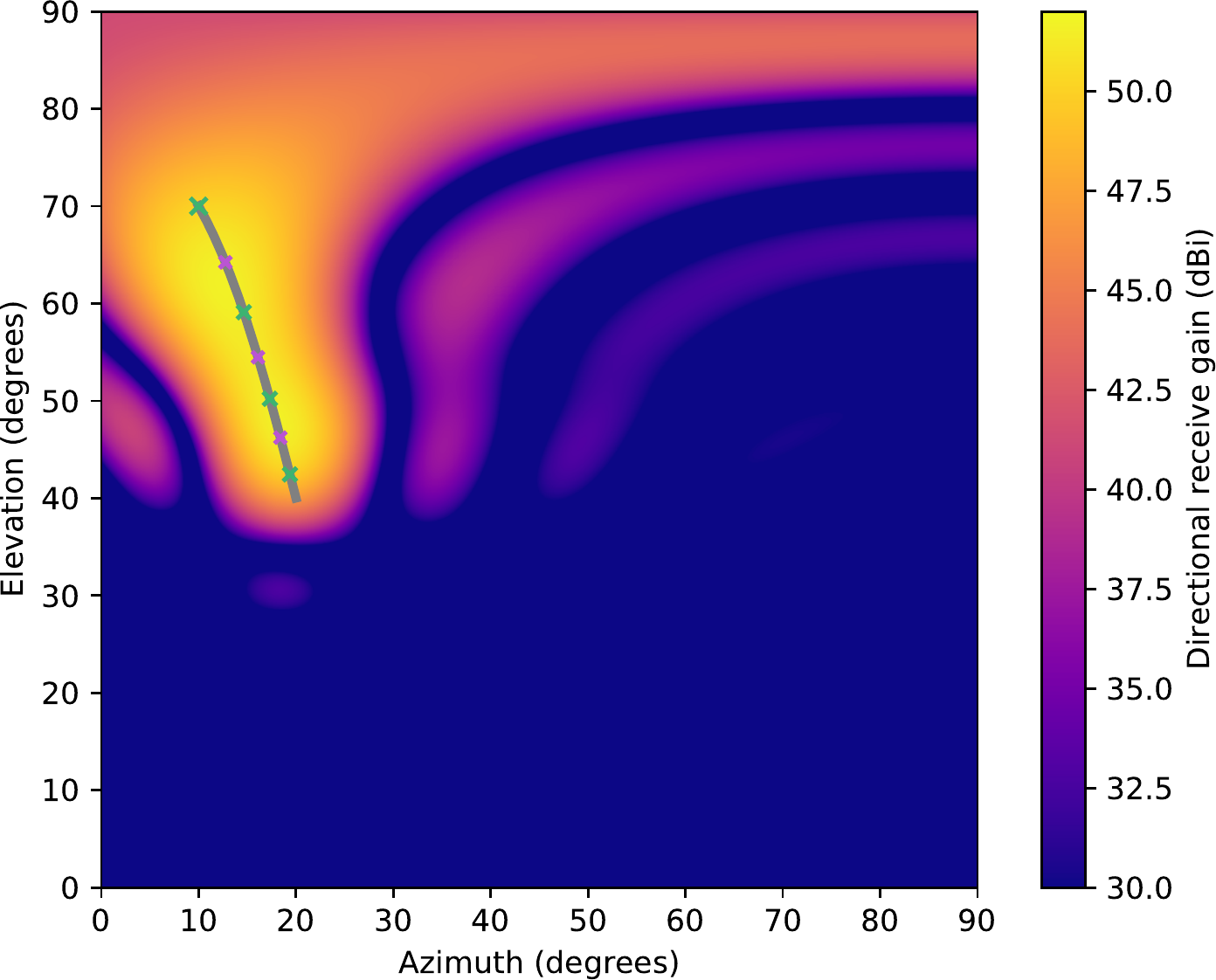}}
    \hfill
    \addtocounter{subfigure}{1}
    \subfloat[Trajectory B, Euler angles]{\includegraphics[width=0.32\textwidth]{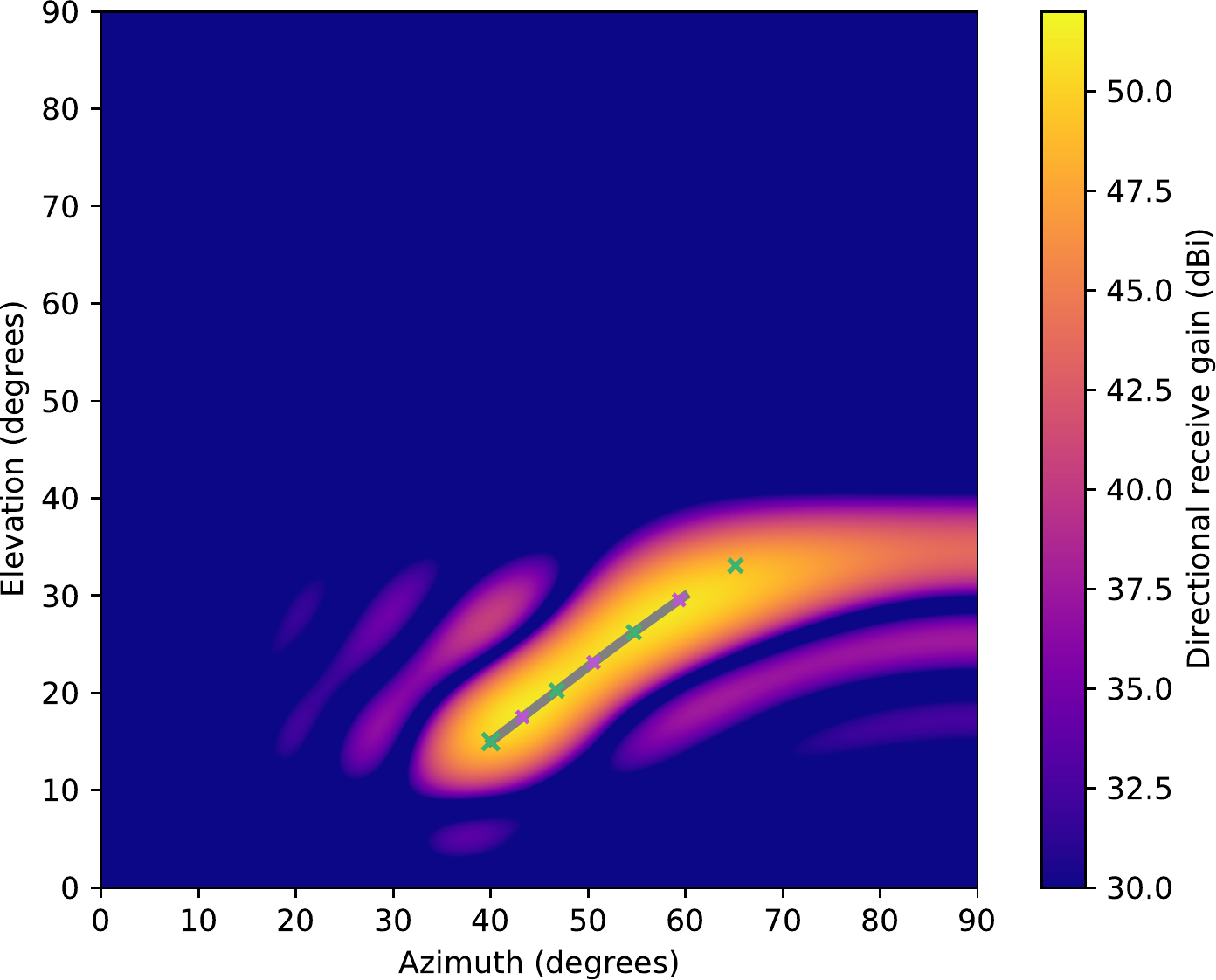}}
    \hfill
    \addtocounter{subfigure}{1}
    \subfloat[Trajectory B, with sub-beam syncing disabled]{\includegraphics[width=0.32\textwidth]{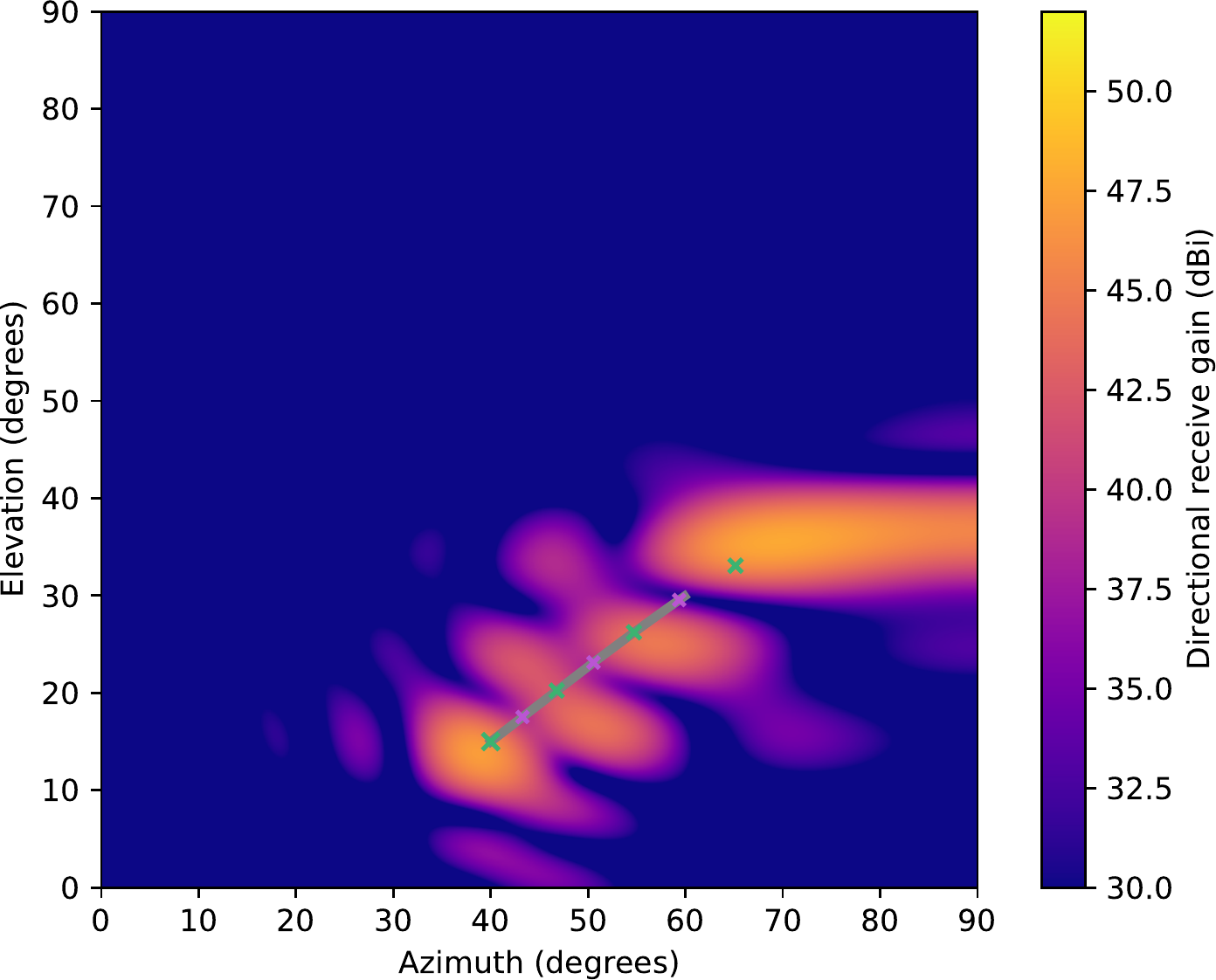} \label{fig:nosmooth}}
    \end{minipage}
    \begin{minipage}{1.0\linewidth}
    \addtocounter{subfigure}{-4}
    \subfloat[Trajectory A, UV-coordinates]{\includegraphics[width=0.32\textwidth]{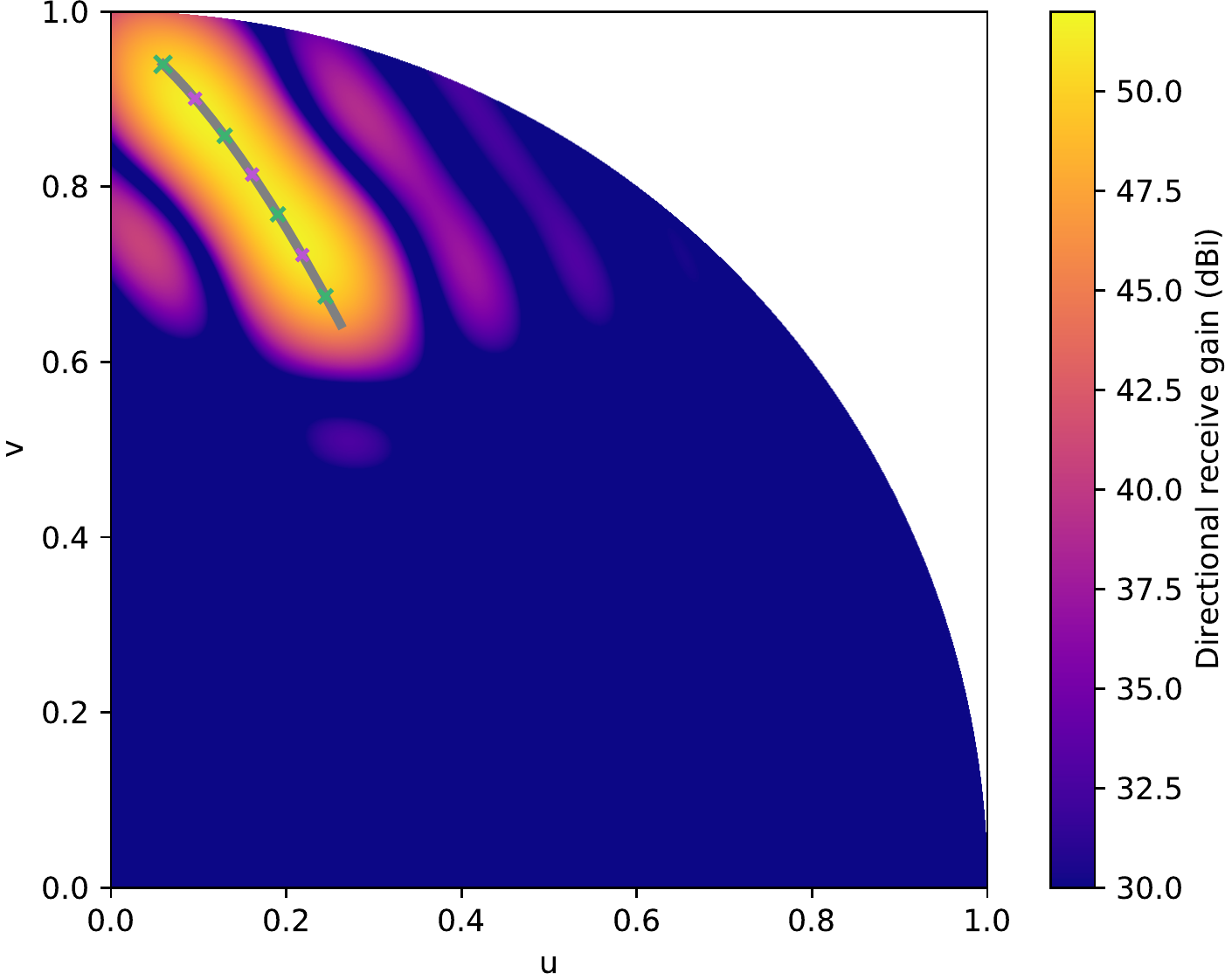}}
    \hfill
    \addtocounter{subfigure}{1}
    \subfloat[Trajectory B, UV-coordinates]{\includegraphics[width=0.32\textwidth]{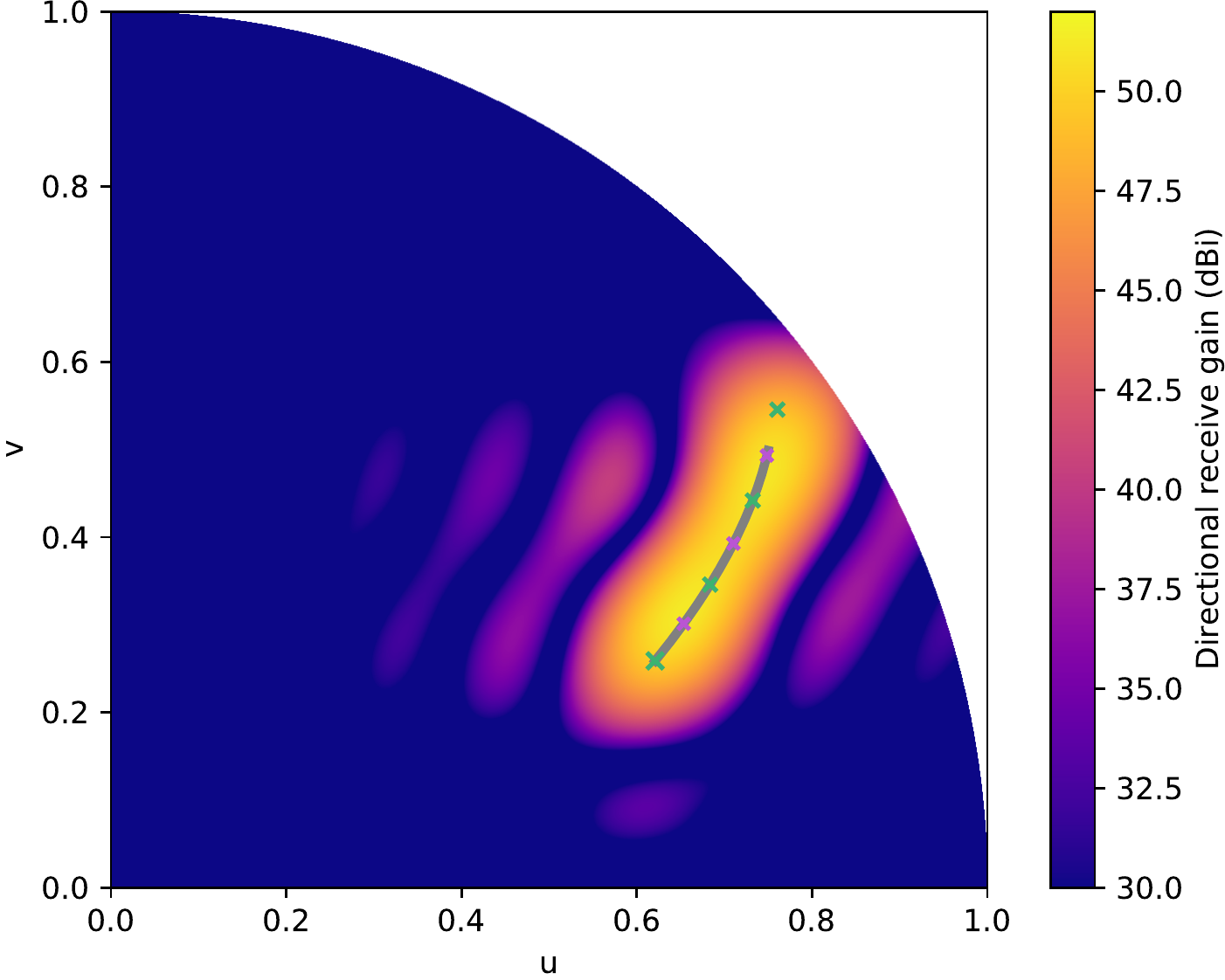}}
    \hfill
    \addtocounter{subfigure}{1}
    \subfloat[Trajectory B, with delayed first sub-beam]{\includegraphics[width=0.32\textwidth]{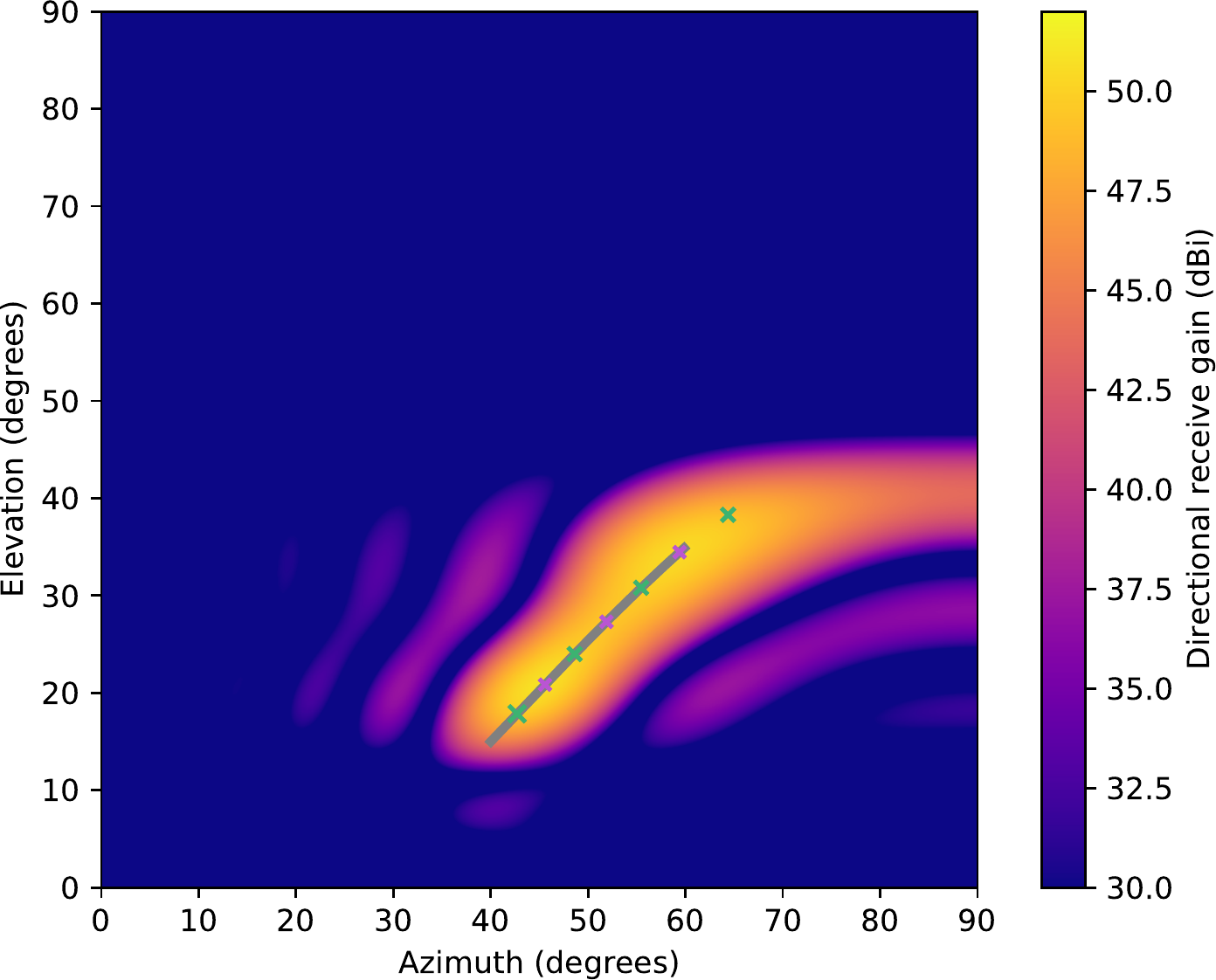} \label{fig:latefirst}} 
    \end{minipage}

    \caption{Directional receive gain using coVRage for two different trajectories. Gains under \SI{30}{\dBi} shown as \SI{30}{\dBi}. Final two images show the effect of disabling features of coVRage. Green and purple crosses indicate sub-beams' focus points and overlap points, respectively.}
    \label{fig:coverages}
\end{figure*}
To calculate the sub-\gls{AWV} $\mathbf{w}_i$ of the $i$-th sub-array, \eqref{eq:beamform} still applies, with $x$ and $y$ being the element indices within the sub-array. To construct the full \gls{AWV}, we first introduce helper functions $f_{i}(x,y)$ and $f_{c}(x,y)$, which map array-wide element coordinates $(x,y)$ to the index of the sub-array said element is assigned to, and to the coordinates within that sub-array, respectively. The elements of the full \gls{AWV} are then
\begin{equation}
    [\mathbf{w}]_{x,y} = \frac{[\mathbf{a}]_{f_{i}(x,y)}}{[\mathbf{w}_{f_{i}(x,y)}]_{f_{c}(x,y)}}
\end{equation}
where $\mathbf{a}$ contains sub-array-level phase shifts, detailed in the following subsection. The first two functions in Algorithm~\ref{alg:bf} summarize this step.
\subsubsection{Sub-Beam Syncing}
Once the sub-array layout is determined and each sub-array is aimed properly, the remaining step is to synchronize the sub-beams, eliminating destructive interference between sub-arrays along the trajectory. As global optimisation at this level is challenging and expensive, we apply a heuristic inspired by previous work on analog sub-arrays~\cite{FlexibleCoverage}. Specifically, we minimize destructive interference between adjacent sub-beams where it is expected to be the most impactful. In this case, this is the point along the trajectory equidistant from the two sub-beams. Sub-beam selection in Section \ref{sec:algo2} was carefully designed to ensure (at least) one sample point of overlap between adjacent beams' coverages. The algorithm iterates through all adjacent sub-beam pairs, determines the phase difference between the two sub-beams at the overlapping point, and applies a uniform additional phase shift to all elements of the second sub-beam, making the two sub-beams phase-aligned at the overlapping sample point.
To determine the phase difference of sub-beams $i$ and $k$ (where $k=i+1$) at point $(u_m,v_m)$, first convert this point to Euler angles $(\phi_m,\theta_m)$ using \eqref{eq:uvtoeuler}. Then determine $C^i_R(\phi_m,\theta_m)$ and $C^k_R(\phi_m,\theta_m)$ by applying \eqref{eq:coeff} with the elements of only sub-array $i$ or $k$. Then set the sub-array-level phase shift such that it undoes this phase difference at sub-array $k$:
\begin{equation}
    [\mathbf{a}]_k = e ^ {j (\angle C^i_R(\phi_m,\theta_m) - \angle C^k_R(\phi_m,\theta_m))}
\end{equation}
where $\angle$ denotes the angle (i.e., the phase) of the complex value. For the first sub-beam, there is no phase shift: $[\mathbf{a}]_0 = 1$. The third function in Algorithm~\ref{alg:bf} summarizes this step.
\subsection{Computational Complexity}
As coVRage is designed to run in real-time on an \gls{HMD}, it must be computationally efficient. The entire procedure consists of closed-form expressions. The first and third functions in Algorithm \ref{alg:bf} are of complexity $O(\log t_l)$ with $t_l$ the trajectory length. Considering the limitations of human head movement, $O(1)$ also approximates their complexity. The second function is $O(|P|)$ with $P$ the sampled points on the trajectory. If required, the sampling rate can be reduced to meet any beamforming deadlines. Any calculated sub-beam direction differs from the optimal direction by at most one sampling interval.
\section{Evaluation}\label{sec:eval}
\begin{figure*}[!t]
    \centering
    \begin{minipage}{1.0\linewidth}
    \subfloat[Trajectory A]{\includegraphics[width=0.24\textwidth]{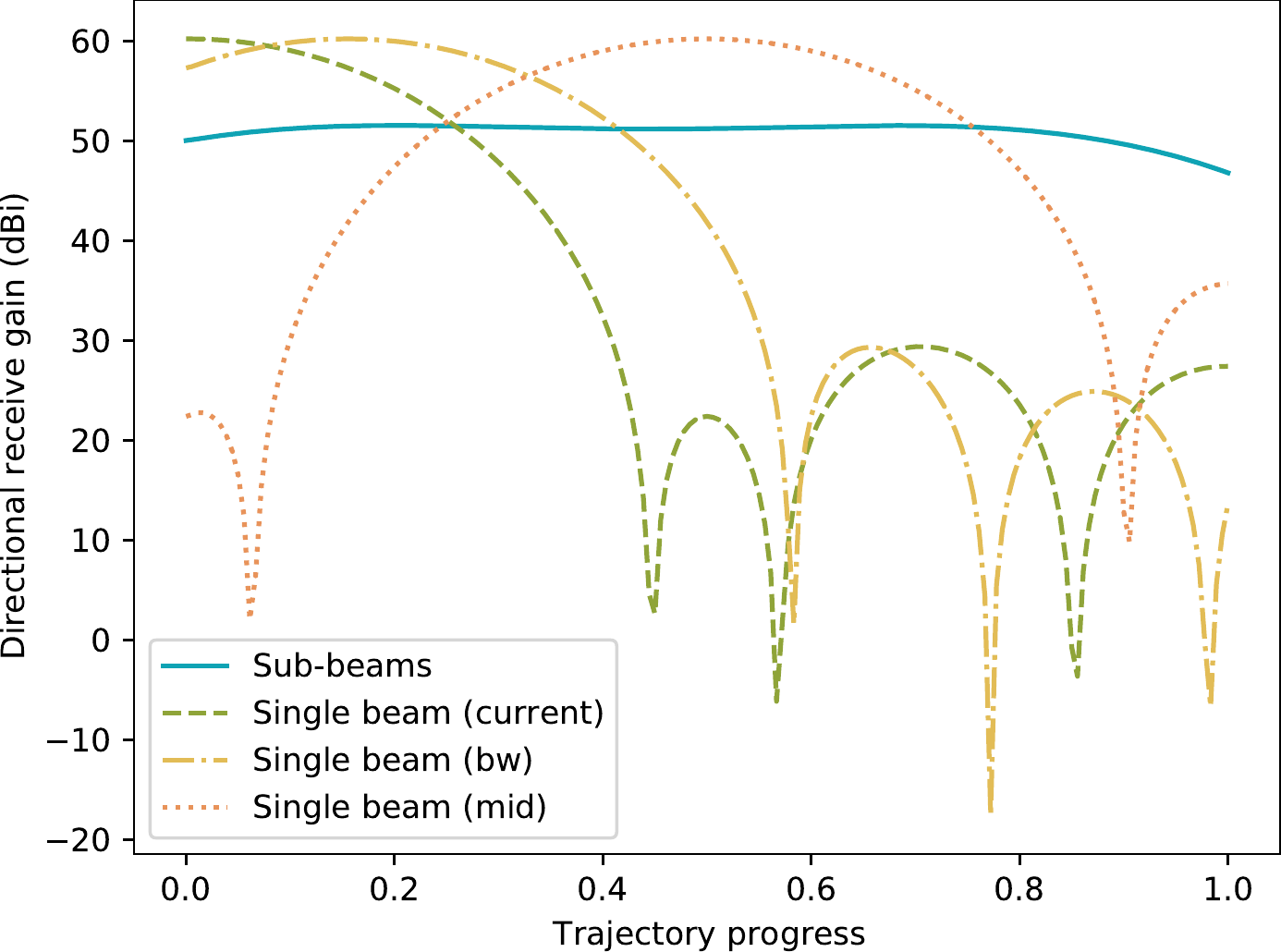} \label{fig:traj1}}
    \hfill
    \subfloat[Trajectory B]{\includegraphics[width=0.24\textwidth]{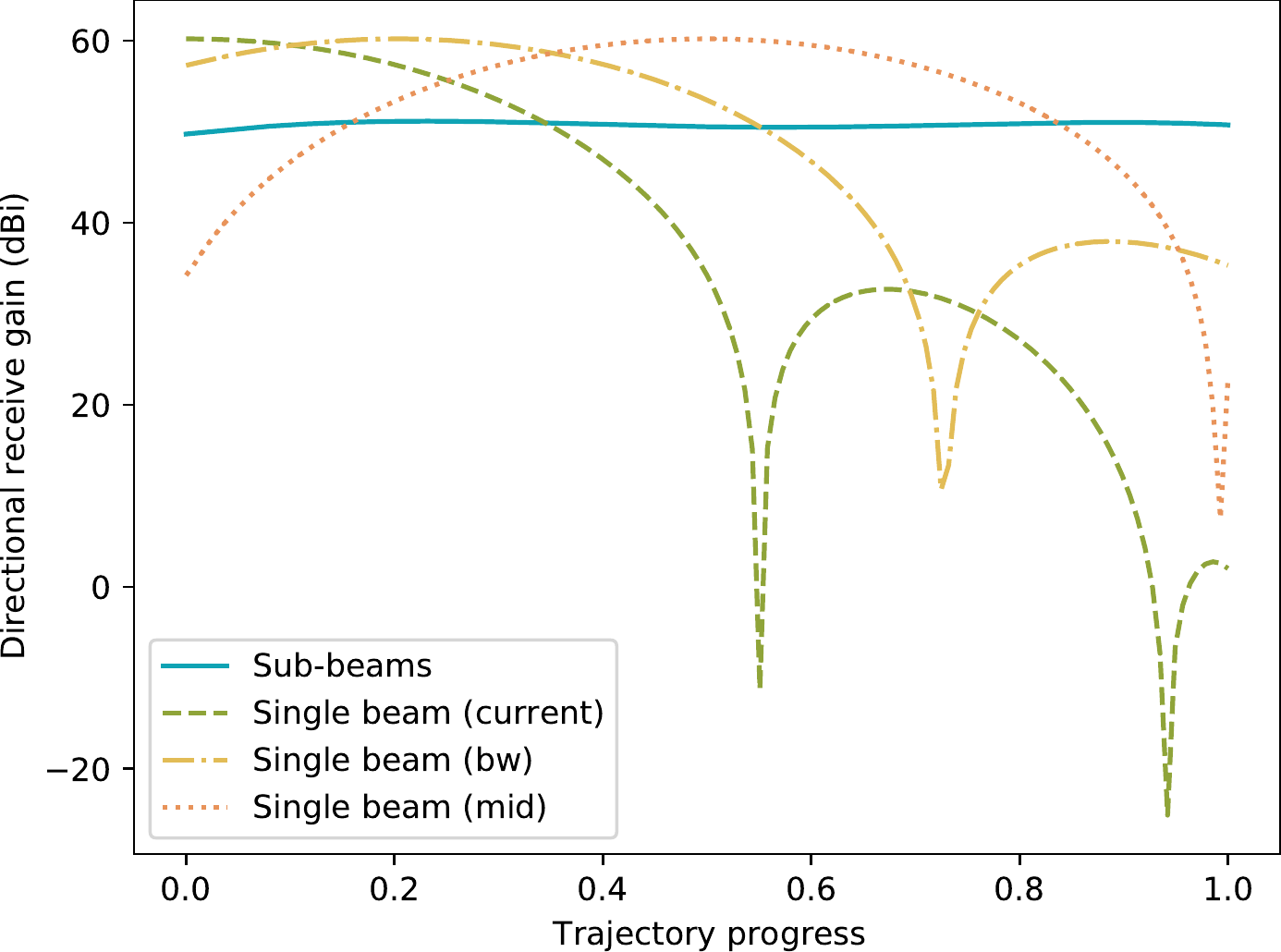} \label{fig:traj2}} 
    \hfill
    \subfloat[Noise penalty for Trajectory A]{\includegraphics[width=0.24\textwidth]{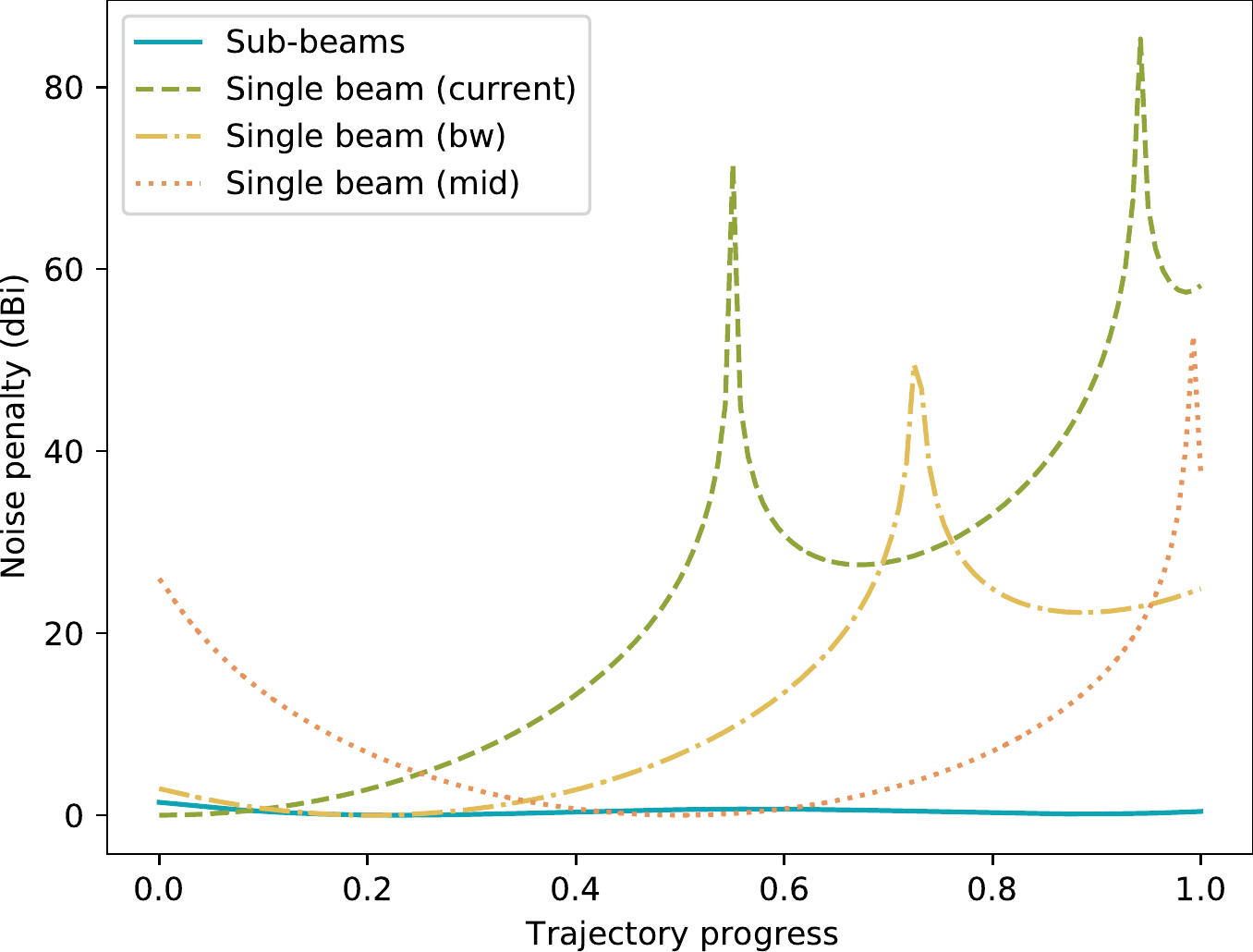} \label{fig:trajNoise}} 
    \hfill
    \subfloat[Noise penalty for Trajectory B]{\includegraphics[width=0.24\textwidth]{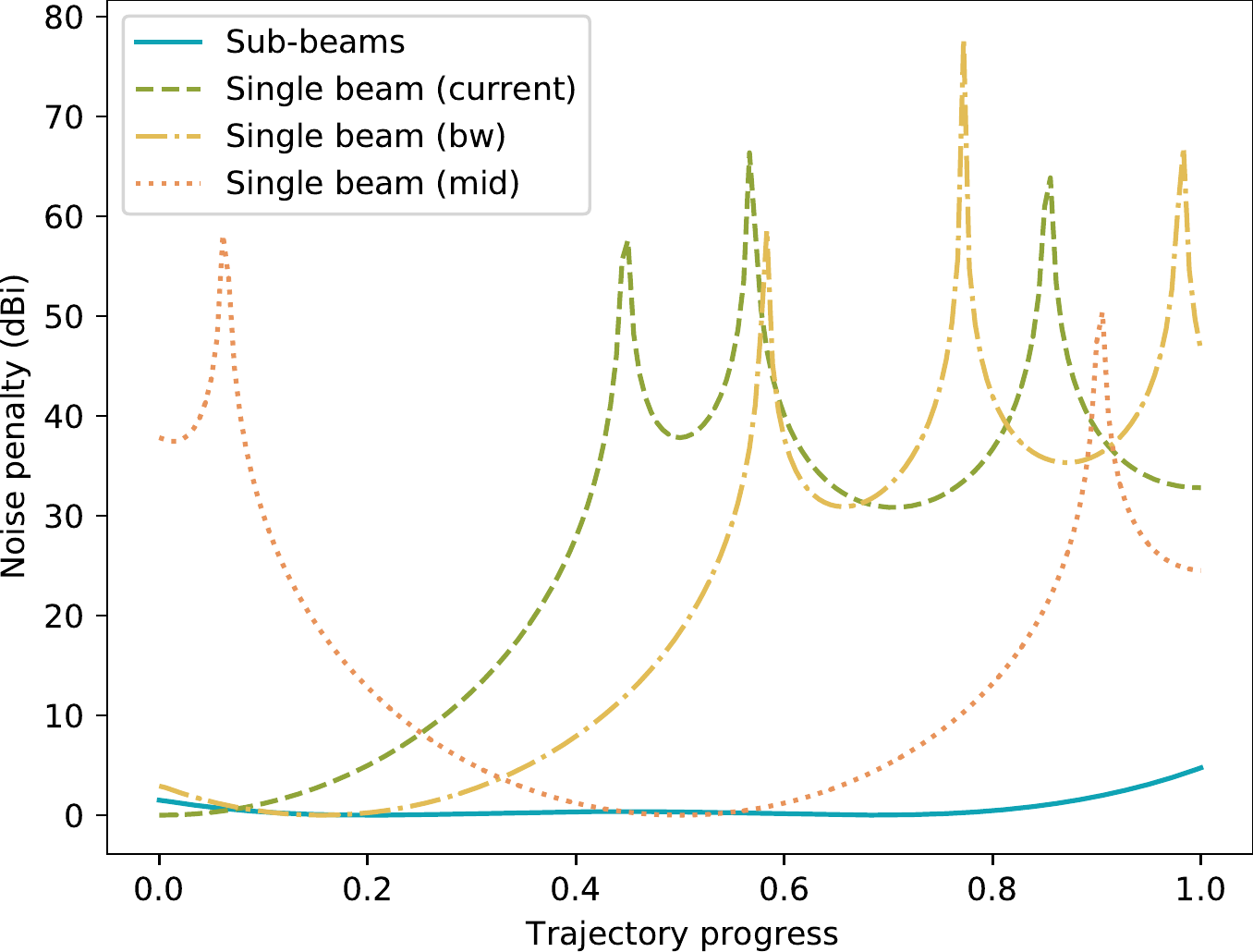} \label{fig:traj2Noise}} 
    \end{minipage}
    \caption{Directional receive gain and noise penalty across the trajectory, using coVRage and single-beam solutions.}
    \label{fig:trajectories}
\end{figure*}

In this section, we simulate coVRage to evaluate how well it performs in the envisioned scenario. First, we assess its performance in its trajectory-covering goal. Then, we analyse the performance within the \gls{VR} application, assessing the impact on attainable datarate using \gls{mmWave} Wi-Fi.\\
To evaluate coVRage, we simulate the \SI{4}{\centi\metre} \SI{60}{\giga\hertz} array, and select two \gls{AP} trajectories requiring all 4 interleaved sub-arrays to be fully covered. Fig.~\ref{fig:coverages} shows the directional receive gain, calculated using \eqref{eq:gain} with both Euler angles and UV-coordinates. For clarity, all gains are raised to at least \SI{30}{\dBi}, and only half the hemisphere is shown. This clearly shows that the gain along the entire trajectory is consistently high. Some deviation from the predicted trajectory is also inherently supported with this beamwidth, without losing excessive energy far away from the trajectory. This provides coVRage with some inherent robustness to prediction errors that may occur with contemporary prediction methods. In trajectory B, extrapolation provided the final sub-beam direction.\\
Next, Fig.~\ref{fig:nosmooth} and \ref{fig:latefirst} illustrate the advantage of some coVRage design decisions. In Fig.~\ref{fig:nosmooth}, sub-beam syncing is disabled, instead using arbitrary, implementation-dependent sub-array-level phase shifts. Overall, the gain is lower, with coverage near the midpoints being especially poor. This indicates that sub-beam syncing is essential to the proper working of the algorithm. In Fig.~\ref{fig:latefirst}, the first sub-beam is not placed at the first trajectory point, but rather at the farthest point whose beam still covers the first point. As there is no sub-beam syncing aimed at optimising gain at this first point, its gain decreases by \SI{7}{\dBi} compared to having a sub-beam pointed directly at it. As this is the actual \gls{AP} direction at the time of beamforming, high coverage for this point is arguably the most important.\\
Next, we evaluate how coVRage compares to steering only a single beam in one specific direction. We consider three possible directions: (1) towards the current \gls{AP} direction, (2) at the farthest trajectory point still covering the current position and (3) halfway along the trajectory. As the single beam uses the full array, with only half the inter-element spacing of the sub-arrays, this beam will be twice as wide as any sub-beam. Using the two trajectories from Fig.~\ref{fig:coverages}, we measure the directional receive gain along the entire trajectory using coVRage and the three single-beam approaches. As Fig.~\ref{fig:traj1} and \ref{fig:traj2} show, the algorithm's coverage of the trajectory is very consistent. In trajectory A, the gain range is \SI{4.75}{\dBi}, largely due to a decrease at the the end of the trajectory. With trajectory B, the final sub-beam is aimed beyond the final trajectory point, meaning coverage remains very stable throughout, with a range of only \SI{1.4}{\dBi}. Higher coverage at the end of the trajectory could be enforced by requiring a final sub-beam beyond the trajectory. 
The single-beam approaches, as expected, manage to outperform coVRage at and around their steering direction. Away from that direction, however, gain reduces quickly, in contrast to coVRage.\\
In analyzing the impact on \gls{SNR}, both the receive gain in the \gls{AoA} direction and the maximum receive gain are of importance. The former determines the intensity of the intended signal, while the latter influences that of the noise, assuming it is isotropic. Therefore, we quantify the approximate impact of gain fluctuations throughout the trajectory using a penalty
\begin{equation}
    N(\phi_{AoA}, \theta_{AoA}) = \max_{\phi,\theta}(G_R(\phi,\theta)) - G_R(\phi_{AoA},\theta_{AoA})
\end{equation}
for some \gls{AoA}. This decibel-scale term can be subtracted from the \gls{SNR} directly, and therefore represents the \gls{SNR} loss caused by high directional gain away from the \gls{AoA} (but possibly elsewhere along the trajectory). As long as the maximum directional gain lies along the trajectory, the gain fluctuation along the trajectory also sets an upper bound to this noise penalty. Fig.~\ref{fig:trajNoise} and \ref{fig:traj2Noise} show the penalty for the two trajectories under consideration. To assess this penalty's impact on performance, we rely on the IEEE 802.11ad standard's minimum received signal intensity for each \gls{MCS}~\cite{EnergyEfficientVR,standard}. From the highest to lowest non-control \gls{MCS}, offering \SI{4620}{Mbps} and \SI{385}{Mbps} throughput respectively, the required intensity drops by \SI{15}{\dB}. As such, when using any single-beam solution, even if the maximum \gls{SNR} is an exceptionally high \SI{25}{\dB} above the requirement for maximum \gls{MCS}, it will drop so low along the trajectory that the datarate reverts to a control-level \SI{27.5}{Mbps} or connectivity is even lost altogether. Either halts delivery of video content to the \gls{HMD} and is extremely disruptive to the user experience. Ignoring the coverage reduction at the end of trajectory A, a maximum \gls{SNR} of just \SI{1.5}{\dB} above the required \gls{SNR} of the highest \gls{MCS} will suffice to maintain the maximum datarate throughout the trajectory. This was previously shown to be sufficient to serve multiple 4K \glspl{HMD} at \SI{120}{\hertz} and a transmission latency under \SI{1}{\milli\second}~\cite{struye2020towards}. Hence, our solution can, in contrast to single-beam solutions, support truly wireless contemporary immersive VR setups.
\section{Conclusion}\label{sec:conclusion}
In this paper, we presented coVRage, the first beamforming algorithm designed specifically for \gls{HMD}-side beamforming with mobile \gls{VR}, where uninterrupted reception even during fast head rotations is crucial for maintaining user experience. Using the \gls{HMD}'s built-in orientation detection capabilities, a predictor can estimate how the \gls{AoA} of incoming wireless video data will change in the near future. By subdividing the phased array into sub-arrays and aiming each sub-array's beam at a different point along the predicted trajectory, coVRage is able to guarantee uninterrupted coverage along the full trajectory, at a very stable signal strength. Simulations using a simple channel model show that coVRage can design beams with a signal strength variation of only a few decibels. A single-beam solution is shown to instead vary by tens of decibels, enough to decimate the attainable throughput, therefore causing a substantial negative impact on, or even fully impairing the end-user’s experience. In future work, we will further investigate capabilities with different array configurations and frequencies. Furthermore, we will quantify the impact of prediction errors and of residual destructive interference between sub-beams, and if needed harden the algorithm against this. Finally, we will combine coVRage with specific trajectory predictors and \gls{AP}-side beamforming to evaluate the performance of an end-to-end system.
\section*{Acknowledgment}
The work of Jakob Struye was supported by the Research Foundation - Flanders (FWO): PhD Fellowship 1SB0719N. The work of Filip Lemic was supported by the EU Marie Skłodowska- Curie Actions Individual Fellowships (MSCA-IF) project Scalable Localization-enabled In-body Terahertz Nanonetwork (SCaLeITN), grant nr. 893760. In addition, this work received support from the University of Antwerp's University Research Fund (BOF).
\bibliographystyle{IEEEtran}
\bibliography{IEEEabrv,bibliography}

\begin{thebibliography}{10}
\providecommand{\url}[1]{#1}
\csname url@samestyle\endcsname
\providecommand{\newblock}{\relax}
\providecommand{\bibinfo}[2]{#2}
\providecommand{\BIBentrySTDinterwordspacing}{\spaceskip=0pt\relax}
\providecommand{\BIBentryALTinterwordstretchfactor}{4}
\providecommand{\BIBentryALTinterwordspacing}{\spaceskip=\fontdimen2\font plus
\BIBentryALTinterwordstretchfactor\fontdimen3\font minus
  \fontdimen4\font\relax}
\providecommand{\BIBforeignlanguage}[2]{{%
\expandafter\ifx\csname l@#1\endcsname\relax
\typeout{** WARNING: IEEEtran.bst: No hyphenation pattern has been}%
\typeout{** loaded for the language `#1'. Using the pattern for}%
\typeout{** the default language instead.}%
\else
\language=\csname l@#1\endcsname
\fi
#2}}
\providecommand{\BIBdecl}{\relax}
\BIBdecl

\bibitem{VRapplications1}
L.~P. Berg and J.~M. Vance, ``Industry use of virtual reality in product design
  and manufacturing: a survey,'' \emph{Virtual reality}, vol.~21, no.~1, pp.
  1--17, 2017.

\bibitem{VRapplications2}
J.~Radianti, T.~A. Majchrzak, J.~Fromm, and I.~Wohlgenannt, ``A systematic
  review of immersive virtual reality applications for higher education: Design
  elements, lessons learned, and research agenda,'' \emph{Computers \&
  Education}, vol. 147, p. 103778, 2020.

\bibitem{VRapplications3}
L.~Li, F.~Yu, D.~Shi, J.~Shi, Z.~Tian, J.~Yang, X.~Wang, and Q.~Jiang,
  ``Application of virtual reality technology in clinical medicine,''
  \emph{American journal of translational research}, vol.~9, no.~9, p. 3867,
  2017.

\bibitem{VRChallenges}
M.~S. {Elbamby}, C.~{Perfecto}, M.~{Bennis}, and K.~{Doppler}, ``Toward
  low-latency and ultra-reliable virtual reality,'' \emph{IEEE Network},
  vol.~32, no.~2, pp. 78--84, 2018.

\bibitem{PerasoVR}
W.~Na, N.-N. Dao, J.~Kim, E.-S. Ryu, and S.~Cho, ``Simulation and measurement:
  Feasibility study of tactile internet applications for mmwave virtual
  reality,'' \emph{ETRI Journal}, vol.~42, no.~2, pp. 163--174, 2020.

\bibitem{VrMecFallback}
Y.~{Liu}, J.~{Liu}, A.~{Argyriou}, and S.~{Ci}, ``Mec-assisted panoramic vr
  video streaming over millimeter wave mobile networks,'' \emph{IEEE
  Transactions on Multimedia}, vol.~21, no.~5, pp. 1302--1316, 2019.

\bibitem{Fundamentals}
D.~Tse and P.~Viswanath, \emph{Fundamentals of Wireless Communication}.\hskip
  1em plus 0.5em minus 0.4em\relax USA: Cambridge University Press, 2005.

\bibitem{MoVR}
O.~Abari, D.~Bharadia, A.~Duffield, and D.~Katabi, ``Enabling high-quality
  untethered virtual reality,'' in \emph{14th {USENIX} Symposium on Networked
  Systems Design and Implementation ({NSDI} 17)}, Boston, MA, Mar. 2017, pp.
  531--544.

\bibitem{OScan}
A.~{Zhou}, L.~{Wu}, S.~{Xu}, H.~{Ma}, T.~{Wei}, and X.~{Zhang}, ``Following the
  shadow: Agile 3-d beam-steering for 60 ghz wireless networks,'' in \emph{IEEE
  INFOCOM 2018 - IEEE Conference on Computer Communications}, 2018, pp.
  2375--2383.

\bibitem{RotationDataset1}
X.~Corbillon, F.~De~Simone, and G.~Simon, ``360-degree video head movement
  dataset,'' in \emph{Proceedings of the 8th ACM on Multimedia Systems
  Conference}, 2017, p. 199–204.

\bibitem{RotationDataset2}
S.~Fremerey, A.~Singla, K.~Meseberg, and A.~Raake, ``Avtrack360: An open
  dataset and software recording people's head rotations watching 360° videos
  on an hmd,'' in \emph{Proceedings of the 9th ACM Multimedia Systems
  Conference}, 2018, p. 403–408.

\bibitem{ShimuraInterleaved}
T.~{Shimura}, T.~{Ohshima}, H.~{Ashida}, S.~{Ishikawa}, S.~{Fujio}, A.~{Honda},
  Z.~{Li}, K.~{Nishikawa}, C.~{Kojima}, K.~{Ozaki}, M.~{Shimizu}, and
  Y.~{Ohashi}, ``Millimeter-wave tx phased array with phase adjusting function
  between transmitters for hybrid beamforming with interleaved subarrays,'' in
  \emph{2016 46th European Microwave Conference (EuMC)}, 2016, pp. 1572--1575.

\bibitem{subarr}
R.~J. {Mailloux}, ``Subarray technology for large scanning arrays,'' in
  \emph{The Second European Conference on Antennas and Propagation, EuCAP
  2007}, 2007, pp. 1--6.

\bibitem{Multimode}
O.~{El Ayach}, R.~W. {Heath}, S.~{Rajagopal}, and Z.~{Pi}, ``Multimode
  precoding in millimeter wave mimo transmitters with multiple antenna
  sub-arrays,'' in \emph{2013 IEEE Global Communications Conference
  (GLOBECOM)}, 2013, pp. 3476--3480.

\bibitem{EnergyEfficient}
X.~{Gao}, L.~{Dai}, S.~{Han}, C.~{I}, and R.~W. {Heath}, ``Energy-efficient
  hybrid analog and digital precoding for mmwave mimo systems with large
  antenna arrays,'' \emph{IEEE Journal on Selected Areas in Communications},
  vol.~34, no.~4, pp. 998--1009, 2016.

\bibitem{ChannelEst}
Y.~{Zhang}, Y.~{Huo}, D.~{Wang}, X.~{Dong}, and X.~{You}, ``Channel estimation
  and hybrid precoding for distributed phased arrays based mimo wireless
  communications,'' \emph{IEEE Transactions on Vehicular Technology}, vol.~69,
  no.~11, pp. 12\,921--12\,937, 2020.

\bibitem{interleaved5G}
{Wenyao Zhai}, V.~{Miraftab}, M.~{Repeta}, D.~{Wessel}, and {Wen Tong},
  ``Dual-band millimeter-wave interleaved antenna array exploiting low-cost pcb
  technology for high speed 5g communication,'' in \emph{2016 IEEE MTT-S
  International Microwave Symposium (IMS)}, 2016, pp. 1--4.

\bibitem{LocalizedvsInterleaved}
J.~A. {Zhang}, X.~{Huang}, V.~{Dyadyuk}, and Y.~J. {Guo}, ``Massive hybrid
  antenna array for millimeter-wave cellular communications,'' \emph{IEEE
  Wireless Communications}, vol.~22, no.~1, pp. 79--87, 2015.

\bibitem{HybridArray_Huang}
X.~{Huang}, Y.~J. {Guo}, and J.~D. {Bunton}, ``A hybrid adaptive antenna
  array,'' \emph{IEEE Transactions on Wireless Communications}, vol.~9, no.~5,
  pp. 1770--1779, May 2010.

\bibitem{HybridAdaptive}
Y.~J. {Guo}, X.~{Huang}, and V.~{Dyadyuk}, ``A hybrid adaptive antenna array
  for long-range mm-wave communications,'' \emph{IEEE Antennas and Propagation
  Magazine}, vol.~54, no.~2, pp. 271--282, 2012.

\bibitem{HybridSDMA}
S.~{Fujio}, C.~{Kojima}, T.~{Shimura}, K.~{Nishikawa}, K.~{Ozaki}, Z.~{Li},
  A.~{Honda}, S.~{Ishikawa}, T.~{Ohshima}, H.~{Ashida}, M.~{Shimizu}, and
  Y.~{Ohashi}, ``Robust beamforming method for sdma with interleaved subarray
  hybrid beamforming,'' in \emph{2016 IEEE 27th Annual International Symposium
  on Personal, Indoor, and Mobile Radio Communications (PIMRC)}, 2016, pp.
  1--5.

\bibitem{HybridMultiplex}
M.~{Shimizu}, A.~{Honda}, S.~{Ishikawa}, K.~{Ozaki}, S.~{Fujio},
  K.~{Nishikawa}, L.~{Zhengyi}, C.~{Kojima}, T.~{Shimura}, H.~{Ashida},
  T.~{Ohshima}, Y.~{Ohashi}, and M.~{Yoshida}, ``Millimeter-wave beam
  multiplexing method using hybrid beamforming,'' in \emph{2016 IEEE 27th
  Annual International Symposium on Personal, Indoor, and Mobile Radio
  Communications (PIMRC)}, 2016, pp. 1--6.

\bibitem{LowComplexMulticast}
Z.~{Li}, C.~{Qi}, and G.~Y. {Li}, ``Low-complexity multicast beamforming for
  millimeter wave communications,'' \emph{IEEE Transactions on Vehicular
  Technology}, vol.~69, no.~10, pp. 12\,317--12\,320, 2020.

\bibitem{AnalogBeamforming}
H.~{Li}, Z.~{Wang}, M.~{Li}, and W.~{Kellerer}, ``Efficient analog beamforming
  with dynamic subarrays for mmwave mu-miso systems,'' in \emph{2019 IEEE 89th
  Vehicular Technology Conference}, 2019, pp. 1--5.

\bibitem{HybridHierarchical}
C.~{Lin}, G.~Y. {Li}, and L.~{Wang}, ``Subarray-based coordinated beamforming
  training for mmwave and sub-thz communications,'' \emph{IEEE Journal on
  Selected Areas in Communications}, vol.~35, no.~9, pp. 2115--2126, 2017.

\bibitem{PartiallyHybridCodebook}
Z.~{Xiao}, P.~{Xia}, and X.~{Xia}, ``Codebook design for millimeter-wave
  channel estimation with hybrid precoding structure,'' \emph{IEEE Transactions
  on Wireless Communications}, vol.~16, no.~1, pp. 141--153, 2017.

\bibitem{VirtualHierarchicalOld}
S.~{Hur}, T.~{Kim}, D.~J. {Love}, J.~V. {Krogmeier}, T.~A. {Thomas}, and
  A.~{Ghosh}, ``Millimeter wave beamforming for wireless backhaul and access in
  small cell networks,'' \emph{IEEE Transactions on Communications}, vol.~61,
  no.~10, pp. 4391--4403, 2013.

\bibitem{VirtualHierarchical}
Z.~{Xiao}, T.~{He}, P.~{Xia}, and X.~{Xia}, ``Hierarchical codebook design for
  beamforming training in millimeter-wave communication,'' \emph{IEEE
  Transactions on Wireless Communications}, vol.~15, no.~5, pp. 3380--3392,
  2016.

\bibitem{AnalogHierarchicalMimo}
Y.~{Sun} and C.~{Qi}, ``Analog beamforming and combining based on codebook in
  millimeter wave massive mimo communications,'' in \emph{2017 IEEE Global
  Communications Conference (GLOBECOM)}, 2017.

\bibitem{AnalogMultiuser}
H.~{Ju}, Y.~{Long}, X.~{Fang}, R.~{He}, and L.~{Jiao}, ``Systematic beam
  management in mmwave networks: Tradeoff among beam coverage, link budget, and
  interference control,'' \emph{IEEE Transactions on Vehicular Technology}, pp.
  1--1, 2020.

\bibitem{FlexibleCoverage}
L.~{Zhu}, J.~{Zhang}, Z.~{Xiao}, X.~{Cao}, D.~O. {Wu}, and X.~{Xia}, ``3-d
  beamforming for flexible coverage in millimeter-wave uav communications,''
  \emph{IEEE Wireless Communications Letters}, vol.~8, no.~3, pp. 837--840,
  2019.

\bibitem{cotsMMVR}
R.~Zhong, M.~Wang, Z.~Chen, L.~Liu, Y.~Liu, J.~Zhang, L.~Zhang, and
  T.~Moscibroda, ``On building a programmable wireless high-quality virtual
  reality system using commodity hardware,'' in \emph{Proceedings of the 8th
  Asia-Pacific Workshop on Systems}, 2017.

\bibitem{renderingVR}
L.~Liu, R.~Zhong, W.~Zhang, Y.~Liu, J.~Zhang, L.~Zhang, and M.~Gruteser,
  ``Cutting the cord: Designing a high-quality untethered vr system with low
  latency remote rendering,'' in \emph{Proceedings of the 16th Annual
  International Conference on Mobile Systems, Applications, and Services},
  2018, p. 68–80.

\bibitem{OffloadingVR}
T.~T. {Le}, D.~V. {Nguyen}, and E.~{Ryu}, ``Computing offloading over mmwave
  for mobile vr: Make 360 video streaming alive,'' \emph{IEEE Access}, vol.~6,
  pp. 66\,576--66\,589, 2018.

\bibitem{struye2020towards}
J.~Struye, F.~Lemic, and J.~Famaey, ``Towards ultra-low-latency mmwave wi-fi
  for multi-user interactive virtual reality,'' \emph{2020 IEEE Global
  Communications Conference (GLOBECOM)}, pp. 1--6, 2020.

\bibitem{EnergyEfficientVR}
J.~Kim, J.-J. Lee, and W.~Lee, ``Strategic control of 60 ghz millimeter-wave
  high-speed wireless links for distributed virtual reality platforms,''
  \emph{Mobile Information Systems}, vol. 2017, 2017.

\bibitem{ProactiveFallback}
S.~Kim and J.-H. Yun, ``Motion-aware interplay between wigig and wifi for
  wireless virtual reality,'' \emph{Sensors}, vol.~20, no.~23, p. 6782, 2020.

\bibitem{ProactiveTransmission}
C.~{Perfecto}, M.~S. {Elbamby}, J.~D. {Ser}, and M.~{Bennis}, ``Taming the
  latency in multi-user vr 360°: A qoe-aware deep learning-aided multicast
  framework,'' \emph{IEEE Transactions on Communications}, vol.~68, no.~4, pp.
  2491--2508, 2020.

\bibitem{Pia}
T.~Wei and X.~Zhang, ``Pose information assisted 60 ghz networks: Towards
  seamless coverage and mobility support,'' in \emph{Proceedings of the 23rd
  Annual International Conference on Mobile Computing and Networking}, 2017, p.
  42–55.

\bibitem{KF5}
J.~J. {LaViola}, ``A comparison of unscented and extended kalman filtering for
  estimating quaternion motion,'' in \emph{Proceedings of the 2003 American
  Control Conference, 2003.}, vol.~3, 2003, pp. 2435--2440 vol.3.

\bibitem{KF3}
E.~{Kraft}, ``A quaternion-based unscented kalman filter for orientation
  tracking,'' in \emph{Sixth International Conference of Information Fusion,
  2003. Proceedings of the}, vol.~1, 2003, pp. 47--54.

\bibitem{KF4}
A.~{van Rhijn}, R.~{van Liere}, and J.~D. {Mulder}, ``An analysis of
  orientation prediction and filtering methods for vr/ar,'' in \emph{IEEE
  Proceedings. VR 2005. Virtual Reality, 2005.}, 2005, pp. 67--74.

\bibitem{DeltaQ}
H.~{Himberg} and Y.~{Motai}, ``Head orientation prediction: Delta quaternions
  versus quaternions,'' \emph{IEEE Transactions on Systems, Man, and
  Cybernetics, Part B (Cybernetics)}, vol.~39, no.~6, pp. 1382--1392, 2009.

\bibitem{Viewport1}
Y.~S. {de la Fuente}, G.~S. {Bhullar}, R.~{Skupin}, C.~{Hellge}, and
  T.~{Schierl}, ``Delay impact on mpeg omaf’s tile-based viewport-dependent
  360° video streaming,'' \emph{IEEE Journal on Emerging and Selected Topics
  in Circuits and Systems}, vol.~9, no.~1, pp. 18--28, 2019.

\bibitem{Viewport2}
S.~{Petrangeli}, G.~{Simon}, and V.~{Swaminathan}, ``Trajectory-based viewport
  prediction for 360-degree virtual reality videos,'' in \emph{2018 IEEE
  International Conference on Artificial Intelligence and Virtual Reality
  (AIVR)}, 2018, pp. 157--160.

\bibitem{ViewportDL1}
T.~{Aykut}, J.~{Xu}, and E.~{Steinbach}, ``Realtime 3d 360-degree telepresence
  with deep-learning-based head-motion prediction,'' \emph{IEEE Journal on
  Emerging and Selected Topics in Circuits and Systems}, vol.~9, no.~1, pp.
  231--244, 2019.

\bibitem{ViewportDL2}
T.~Aykut, B.~G{\"u}lezy{\"u}z, B.~Girod, and E.~Steinbach, ``Hsmf-net: Semantic
  viewport prediction for immersive telepresence and on-demand 360-degree
  video,'' \emph{arXiv preprint arXiv:2009.04015}, 2020.

\bibitem{riftAccuracy}
T.~A. Jost, B.~Nelson, and J.~Rylander, ``Quantitative analysis of the oculus
  rift s in controlled movement,'' \emph{Disability and Rehabilitation:
  Assistive Technology}, vol.~0, no.~0, pp. 1--5, 2019, pMID: 31726896.

\bibitem{RiftTracking}
S.~M. {LaValle}, A.~{Yershova}, M.~{Katsev}, and M.~{Antonov}, ``Head tracking
  for the oculus rift,'' in \emph{2014 IEEE International Conference on
  Robotics and Automation (ICRA)}, 2014, pp. 187--194.

\bibitem{phaseBook}
R.~J. Mailloux, \emph{Phased array antenna handbook}.\hskip 1em plus 0.5em
  minus 0.4em\relax Artech house, 2017.

\bibitem{RedirectedWalking}
E.~R. {Bachmann}, E.~{Hodgson}, C.~{Hoffbauer}, and J.~{Messinger},
  ``Multi-user redirected walking and resetting using artificial potential
  fields,'' \emph{IEEE Transactions on Visualization and Computer Graphics},
  vol.~25, no.~5, pp. 2022--2031, 2019.

\bibitem{MOCA}
M.~K. Haider and E.~W. Knightly, ``Mobility resilience and overhead constrained
  adaptation in directional 60 ghz wlans: Protocol design and system
  implementation,'' in \emph{Proceedings of the 17th ACM International
  Symposium on Mobile Ad Hoc Networking and Computing}, 2016, p. 61–70.

\bibitem{ZeroOverhead}
A.~Loch, H.~Assasa, J.~Palacios, J.~Widmer, H.~Suys, and B.~Debaillie, ``Zero
  overhead device tracking in 60 ghz wireless networks using multi-lobe beam
  patterns,'' in \emph{Proceedings of the 13th International Conference on
  Emerging Networking EXperiments and Technologies}, 2017, p. 224–237.

\bibitem{pathloss}
A.~Maltsev, E.~Perahia, R.~Maslennikov, A.~Lomayev, A.~Khoryaev, and
  A.~Sevastyanov, ``Path loss model development for tgad channel models,''
  \emph{IEEE 802.11--09/0553r1}, 2009.

\bibitem{Angles1}
J.~Diebel, ``Representing attitude: Euler angles, unit quaternions, and
  rotation vectors,'' 2006.

\bibitem{Angles2}
M.~D. Shuster \emph{et~al.}, ``A survey of attitude representations,''
  \emph{The Journal of the Astronautical Sciences}, vol.~41, no.~4, pp.
  439--517, 1993.

\bibitem{Angles3}
G.~{Taubin}, ``3d rotations,'' \emph{IEEE Computer Graphics and Applications},
  vol.~31, no.~6, pp. 84--89, 2011.

\bibitem{UVcoords}
W.~H. {Von Aulock}, ``Properties of phased arrays,'' \emph{Proceedings of the
  IRE}, vol.~48, no.~10, pp. 1715--1727, 1960.

\bibitem{Slerp}
K.~Shoemake, ``Animating rotation with quaternion curves,'' \emph{SIGGRAPH
  Comput. Graph.}, vol.~19, no.~3, p. 245–254, Jul. 1985.

\bibitem{standard}
``Ieee standard for information technology—telecommunications and information
  exchange between systems local and metropolitan area networks—specific
  requirements - part 11: Wireless lan medium access control (mac) and physical
  layer (phy) specifications,'' \emph{IEEE Std 802.11-2016 (Revision of IEEE
  Std 802.11-2012)}, pp. 1--3534, 2016.

\end{thebibliography}

\end{document}